\documentclass[11pt]{article}

\usepackage{sectsty}
\usepackage{graphicx}
\usepackage{authblk}
\usepackage{amsmath}
\title{Impact of inclusive jet cross sections with low transverse momenta on the determination of gluon parton distributions in pp collisions}
\author[1]{D. Sunar Cerci\footnote{e-mail: deniz.sunar.cerci@cern.ch}}
\author[1]{S. Cerci}
\author[2]{K. Wichmann}
\affil[1]{Adiyaman University, Faculty of Arts and Sciences, Department of Physics, Turkiye}
\affil[2]{Deutsches Elektronen-Synchrotron DESY, Hamburg, Germany}

\date{}

\begin{document}
\maketitle	



\begin{abstract}
Jet production at hadron colliders provides constraints on the parton distribution functions (PDFs) of the proton, in particular on the gluon distribution. In the present paper, 
the impact on PDF of CMS inclusive differential jet cross sections at center-of-mass energy of $\sqrt s = 8$ TeV for jets with low momentum $p\mathrm{_T}$ and produced in the forward 
direction is investigated at next-to leading order in perturbative quantum chromodynamics (QCD).
The results of the QCD global analysis are compared with theoretical predictions. 
The impact of low-$p\mathrm{_T}$ jet measurements on the determination of the gluon distribution is assessed. 
The inclusion of the discussed measurements adds further constraints on uncertainty of the gluon distribution at large  Bjorken $x > 0.1$, where the low-$p_T$ data have the largest impact.

\end{abstract}

\section{Introduction}
\label{intro}
The  parton distribution functions (PDFs) represent the probability densities 
to find a parton in the proton carrying a momentum fraction $x$ at a squared energy scale and 
 are an essential ingredient for precise knowledge of the structure of the proton. 
 The factorization scale dependence of PDFs is purely perturbative and can be determined by the DGLAP evolution equations~\cite{GRIBOV197178,Lipatov:400357,ALTARELLI1977298}. 
 However, the dependence on the momentum fraction is non-perturbative and it has to be extracted from experimental measurements. 
 Deep inelastic scattering (DIS) data from HERA~\cite{h12015combination} are core of any PDF extraction, 
 covering a big part of the kinematic phase space, with the Bjorken $x$  down to $x <~ 10^{-5}$
    and the negative four-momentum-transfer squared $Q^2$ up to $50000$ GeV$^2$.

With the start of LHC data taking, more and more $pp$ data is included in the global QCD analyses to improve
determination of parton distributions in the proton. 
Jet and $t\bar{t}$ production cross sections can provide new constrain on gluon distribution. 
The production of W or Z bosons at the LHC is used in particular to improve the knowledge of valence distributions.
These data are also indirectly sensitive to the strange quark distribution. 
The direct constrains on the strange sea is coming from the LHC measurements of W+charm cross sections.
All these data are now used in PDF determinations of various collaborations, like
CT18~\cite{CT18}, MSHT20~\cite{Bailey2021}, NNPDF4~\cite{Ball2022}, ABMP~\cite{Alekhin_2017}, JR \cite{Jimenez_Delgado_2014}, 
and CJ \cite{Accardi_2016}. 

The xFitter~\cite{alekhin2015herafitter,bertone2017xfitter} is an open-source platform that
allows to extract PDFs and to assess the impact of new data.  
A variety of theoretical predictions for different processes together with a large number of existing methods for
determining the proton and nuclear PDFs and model parameters (like the strong coupling or quark masses) 
are available in the modular structure of xFitter. 
It is extensively used by the CMS and ATLAS collaborations to study the impact of their data in determination of parton densities.
Some examples are given here~\cite{Aaboud_2017,Aaboud_2017_2,ATLASepWZVjet20,Khachatryan_2017,cmsWc13,cmsJets_2021}.

\section{Analysis setup}
\label{sec:1}

\subsection{Theoretical framework}
\label{ssec:2-1}
In this paper, the xFitter version 2.0.0 is used to estimate 
the impact of the CMS inclusive low-$p\mathrm{_T}$ jet cross section measurement 
at $\sqrt{s} = 8$~TeV~\cite{Khachatryan_2017} on the PDFs and their uncertainties.
A PDF fit at next-to-leading order (NLO) is performed, using the HERA DIS cross 
sections~\cite{h12015combination}, and the 
jet cross sections~\cite{Khachatryan_2017}.  
In xFitter the parton distributions are evolved using the DGLAP equations 
at different orders in QCD, 
as implemented in the QCDNUM program~\cite{Botje_2011}.  The heavy-quark contributions are treated 
using the \mbox{generalised-mass} variable-flavour number \mbox{Roberts-Thorne} 
scheme ~\cite{Thorne_1998,Thorne_2006,Thorne_2012}. 
Here the charm and beauty quark masses are model parameters and are fixed to $m_c = 1.46 \pm 0.04$~GeV 
and $m_b = 4.3 \pm 0.1$~GeV, respectively, following the newest estimates from
HERA~\cite{HERAPDF2.0JetsNNLO}. 
The theoretical predictions for the cross sections of jet production are calculated at NLO
by using the NLOJET++ program~\cite{Nagy_2002,Nagy_2003} as implemented into the FASTNLO package~\cite{britzger2012new}. 
The factorisation and renormalisation scales are set to the four-momentum transfer ($Q$) for the DIS 
processes i.e,  $\mu_f =\mu_r = \sqrt{Q^2}$ and to the jet $p\mathrm{_T}$ in case of the CMS jet cross sections. 
The PDF extraction is done using a $\chi^{2}$ minimization method, as implemented in 
MINUIT~\cite{JAMES1975343} framework.
In the results presented here a cut on minimum $Q^2$ of the HERA data, $Q^2_{min} \ge 7.5~\mathrm{GeV^2}$, is required. 
Also all figures presented in this study, excluding Fig.~\ref{fig:1} are done using xFitter.

\subsection{Jet data} 
\label{ssec:2-2}
\begin{figure}[h!tb]
\begin{center} 
 \includegraphics[width=0.5\textwidth]{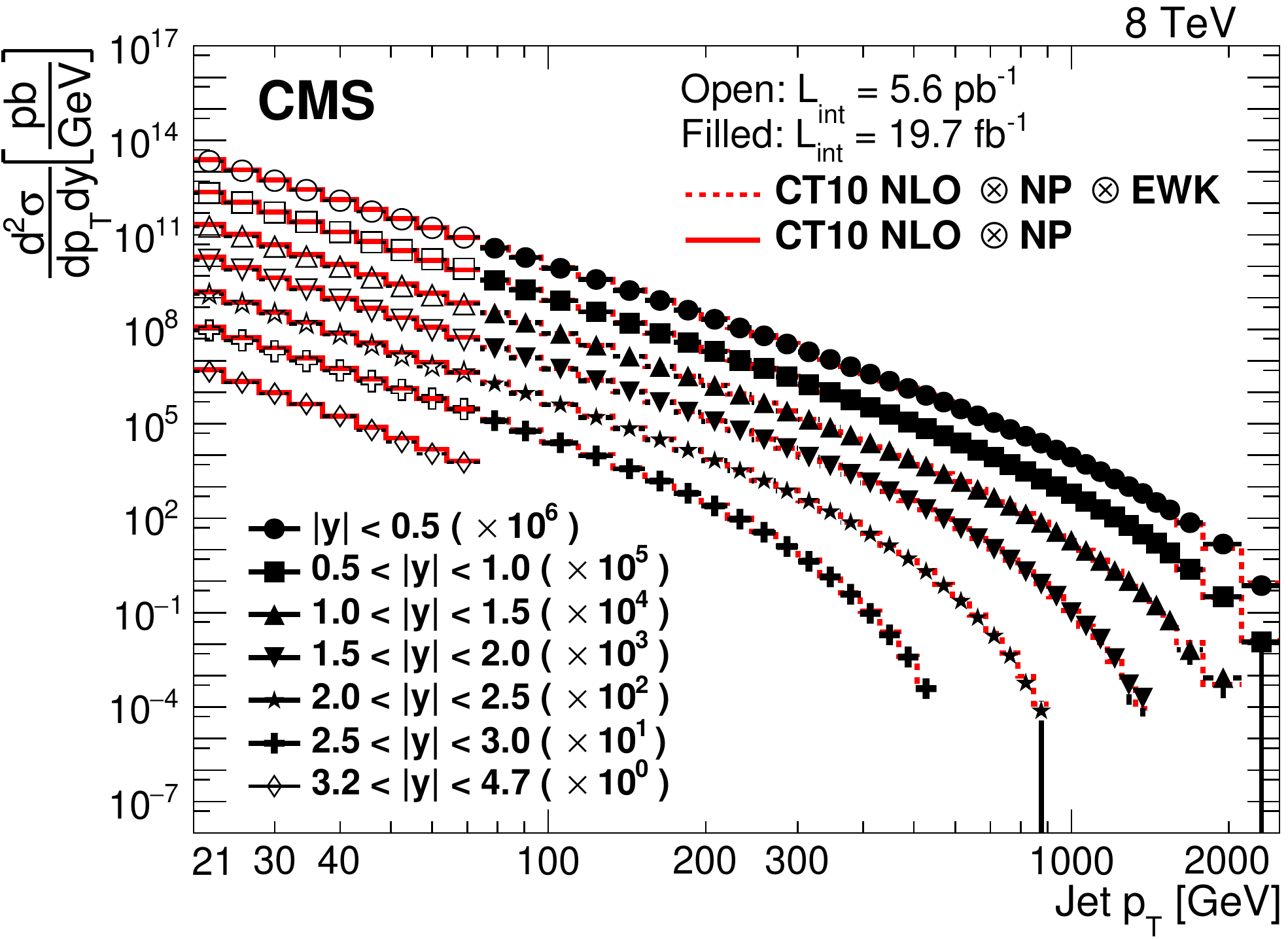}  
  \caption{CMS double-differential inclusive jet cross sections at 8 TeV~\cite{Khachatryan_2017} shown as function of jet $p\mathrm{_T}$. The low-$p\mathrm{_T}$  data are represented by open points and the high-$p\mathrm{_T}$ data by filled points.}
\label{fig:1}
\end{center}
\end{figure}

In this paper the CMS inclusive jet measurement at 8 TeV~\cite{Khachatryan_2017} is considered.
It includes two sets of jet data, one of which is named high-$p\mathrm{_T}$ and the other is low-$p\mathrm{_T}$.
The $p\mathrm{_T}$ ranges considered in  the low-$p\mathrm{_T}$ and 
high-$p\mathrm{_T}$ jet analyses are $21-74$~GeV and $74-2500$~GeV, respectively. 
These double-differential inclusive jet cross sections are shown in Fig.~\ref{fig:1}~\cite{Khachatryan_2017}.
The data included in the high-$p\mathrm{_T}$ jet measurement are collected using six single-jet triggers in 
the high-level trigger system requiring at least one jet in the event with jet 
$p\mathrm{_T} > 40, 80, 140, 200, 260,$ and 320 GeV, while events for 
the low-$p\mathrm{_T}$ jet analysis are selected online in an unbiased way by triggering the low average number of $pp$ interactions per bunch crossing, or pileup. 

The infrared and collinear safe anti-$k\mathrm{_T}$ jet clustering algorithm~\cite{Cacciari2008} is used to reconstruct jets   
with a distance parameter of $R = 0.7$.  
The inputs to the jet clustering algorithm are the four-momentum vectors of particle-flow objects. 
Individual particles (leptons, photons, charged and neutral hadrons) are reconstructed with the particle-flow technique~\cite{cmspf0} which combines the information from several sub-detectors. 

In order to account for various experimental effects, the raw jet $p\mathrm{_T}$ spectra are corrected. 
The reconstructed jet energy is calibrated with correction factors derived using real data, via a $p\mathrm{_T}$-balancing method in dijet and in photon-jet events, 
as well as from MC simulations~\cite{cmspf1,cmspf2}. 
The jet energy scale (JES) correction is applied to the jet four-momentum vector as a multiplicative factor.
It depends on the $\eta$ and $p\mathrm{_T}$ of the jet.
The JES corrections are different for both analyzes, especially in $p\mathrm{_T}$-dependent components. 
This also contributes to fluctuations in the transition between the low- and high-$p\mathrm{_T}$ jet regions. 
Although this causes some discontinuity to be observed in the measured values, both analyzes are generally compatible 
in terms of total experimental uncertainties. 
Since the initial part of the 2012 low pileup data sample was used in the low-$p\mathrm{_T}$ jet analysis, there is no JES time dependence and the pileup-related corrections are negligible. 
Systematic uncertainty sources for the high-$p\mathrm{_T}$ and low-$p\mathrm{_T}$ jet data sets are treated as correlated, 
except for those with the pileup-related corrections and JES time dependence which are assumed to be uncorrelated.

The high-$p\mathrm{_T}$ jet data set has already been used
in a global QCD analysis in the CMS publication~\cite{Khachatryan_2017} to extract parton densities and $\alpha_s(M_Z^2)$.
This analysis illustrated that the high-$p\mathrm{_T}$ jet cross sections provide important constraints 
on the gluon distributions in a new corner of kinematic phase space.

\subsection{Parameterisation}
\label{sec:param}

In the QCD analysis we use the approach of HERAPDF~\cite{h12015combination}, where  
the PDFs of the proton, $xf$, are parameterised at the starting scale 
$\mu^2_{\rm f_{0}} = 1.9$ $\mathrm{GeV^{2}}$ in a general way as
\begin{equation}
 xf(x) = A x^{B} (1-x)^{C} (1 + D x + E x^2)~.
\label{eqn:pdf}
\end{equation}
Here $x$ is the fraction of the proton's momentum taken by the struck parton 
in the infinite momentum frame. 
The parameterisations for the following PDFs are used: the gluon distribution, $xg$, 
the valence quark distributions, $xu_v$, $xd_v$, and 
the $u$-type and $d$-type anti-quark distributions,
$x\bar{U}$, $x\bar{D}$. We assume that $x\bar{U} = x\bar{u}$ and 
$x\bar{D} = x\bar{d} +x\bar{s}$ at the starting scale $\mu^2_{\rm f_{0}}$.
  
The nominal parameterisation used in the studies presented here is
\begin{eqnarray}
\label{eq:xgpar}
xg(x) &=   & A_g x^{B_g} (1-x)^{C_g} - A_g' x^{B_g'} (1-x)^{C_g'}  ,  \\
\label{eq:xuvpar}
xu_v(x) &=  & A_{u_v} x^{B_{u_v}}  (1-x)^{C_{u_v}}\left(1+D_{u_v}x+E_{u_v}x^2 \right) , \\
\label{eq:xdvpar}
xd_v(x) &=  & A_{d_v} x^{B_{d_v}}  (1-x)^{C_{d_v}} , \\
\label{eq:xubarpar}
x\bar{U}(x) &=  & A_{\bar{U}} x^{B_{\bar{U}}} (1-x)^{C_{\bar{U}}}\left(1+D_{\bar{U}}x\right) , \\
\label{eq:xdbarpar}
x\bar{D}(x) &= & A_{\bar{D}} x^{B_{\bar{D}}} (1-x)^{C_{\bar{D}}}\left(1+E_{\bar{D}}x^2\right) .
\end{eqnarray}

As can be seen from Eq.~\ref{eq:xgpar} the gluon distribution, $xg$, is an exception from Eq.~\ref{eqn:pdf}, 
from which an additional term of the form 
$A_g'x^{B_g'}(1-x)^{C_g'}$is subtracted\footnote{In this analysis, 
$C_g'$ is fixed to $C_g' = 25$~\cite{Martin:2009iq}, following~\cite{h12015combination}.}.
$A_{u_v}, A_{d_v} and A_g$ are the normalisation parameters constrained 
by the momentum sum rule and the quark-number sum rules. 
The $B_{\bar{U}}$ and $B_{\bar{D}}$ parameters were set as equal,
$B_{\bar{U}}=B_{\bar{D}}$. 
The strange quark distribution is taken 
as an $x$-independent fraction, $f_s$, of the $d$-type sea, 
$x\bar{s}= f_s x\bar{D}$ at $\mu^2_{\rm f_{0}}$.
Here $f_s=0.4$ and additionally
$A_{\bar{U}}=A_{\bar{D}} (1-f_s)$.

The final parameterisation, so the parameters appearing in Eqs.~\ref{eq:xgpar}--\ref{eq:xdbarpar} 
were selected using the so-called "parameterisation scan"~\cite{h12015combination}.
First fit with all $D$ and $E$ parameters and $A_g'$ set to zero was performed (10 free parameters). 
Then we included in the fit  other parameters one at a time and we looked at the improvement of the  $\chi^2$ of the fits.
We stopped the procedure when no further improvement in  $\chi^2$ was seen.
This gave a $16$-parameter fit.

\subsection{Uncertainties}
\label{SS:3-1}

The way of estimating the experimental, model and parameterisation uncertainties of the PDFs was chosen as in 
the HERAPDF method.
To assess experimental uncertainties the Hessian method based on the $\Delta\chi^2=1$ is used.

The model uncertainty includes variation of the $Q^2_{min}$ cut value, 
the values of the starting scale, the heavy quark masses and $f_s$, 
following the HERAPDF2.0 analysis~\cite{HERAPDF2.0JetsNNLO}. 
The only difference is for the $Q^2_{min}$ cut value, which is larger in the present analysis. 
It was varied between 5 and 10 GeV$^2$.
To obtain the model uncertainty, we added in quadrature, separately for positive and negative deviations,
the differences between the nominal fit and the fits corresponding to the variations described above.

The parameterisation uncertainty is defined as an envelope of the results
obtained by changing the starting scale $\mu^2_{\rm f_{0}}$ to 1.6 and 2.2 $\mathrm{GeV^{2}}$,
and by adding, one at a time, extra $D$ and $E$ parameters in the polynomials of Eqs.~\ref{eq:xgpar}--\ref{eq:xdbarpar}.

The total uncertainty is given by adding in quadrature 
the experimental, the model and the parameterisation uncertainties.

\section{Results}
\label{SS:results}

\subsection{Estimation of $\alpha_s(M_Z^2)$ }
\label{S:alphas}

A simultaneous QCD fit of PDFs and  $\alpha_s(M_Z^2)$ was performed using HERA DIS and CMS jet data.
The experimental, model and parameterisation uncertainties of $\alpha_s(M_Z^2)$ were estimated as described above. 
To evaluate the scale uncertainty on $\alpha_s(M_Z^2)$ the renormalisation and factorisation 
scales were varied up and down by a factor of two and $\alpha_s(M_Z^2)$ estimated from each such fit.
The maximal positive and negative deviations on $\alpha_s(M_Z^2)$ (with exclusion of the two extreme combinations 
of the scales) were taken as the scale uncertainty.
The uncertainties were assumed to be 100\,\% correlated between bins and data sets.\\

For the PDF fit including low-$p\mathrm{_T}$ and high-$p\mathrm{_T}$ jet data at 8 TeV, the measured $\alpha_s(M_Z^2)$ is
\begin{align}
\alpha_s(M_Z^2) = 0.1199 &\pm  0.0011~(exp)~^{+0.0007}_{-0.0004}~(model)~^{+0.0002}_{-0.0005}~(param) 
~^{+0.0032}_{-0.0011}~(scale). \nonumber
\end{align}

We also performed fits with $\alpha_s(M_Z^2)$ as a free parameter for low-$p\mathrm{_T}$ and high-$p\mathrm{_T}$ 
jet data only (using the same parameterisation). The results are summarised below.\\

For the PDF fit including high-$p\mathrm{_T}$ jet data, the measured $\alpha_s(M_Z^2)$ is
\begin{align}
\alpha_s(M_Z^2) = 0.1218 &\pm 0.0021~(exp)~^{+0.0004}_{-0.0006}~(model)~\pm 0.0004~(param)~^{+0.0025}_{-0.0009}~(scale).\nonumber
\end{align}

For the PDF fit including low-$p\mathrm{_T}$ jet data, the measured $\alpha_s(M_Z^2)$ is
\begin{align}
\alpha_s(M_Z^2) = 0.1178 &\pm 0.0029~(exp)~^{+0.0006}_{-0.0020}~(model)~\pm 0.0011~(param)~^{+0.0028}_{-0.0012}~(scale).\nonumber \\
\end{align}

\begin{table*}[h]
\begin{center}
\caption{Total and partial $\chi^2$ for the HERA and CMS data included in the PDF fits.}
\label{tab:1}
\begin{tabular}{c|ccc}
\hline
 & & CMS 8 TeV jet data & \\
             & high-$p\mathrm{_T}$&     low-$p\mathrm{_T}$           & high- + low-$p\mathrm{_T}$ \\
             & $\alpha_s(M_Z^2) = 0.1218$ & $\alpha_s(M_Z^2) = 0.1178$ & $\alpha_s(M_Z^2) = 0.1199$\\
\hline
Total $\chi^2/\rm{NDF}$               & 1410/1207 & 1307/1102  &  1583/1270\\
\hline
  Data sets         & &$\chi^2/\rm{NDP}$    &  \\
\hline                
HERA                 & 1159/1056 & 1141/1056  & 1162/1056 \\
\hline  
high-$p\mathrm{_T}$ data                             &                &       &         \\
 $0.0 < |y| < 0.5$          &    43/35       &       &   46/35  \\ 
 $0.5 < |y| < 1.0$          &    31/34       &       &   40/34 \\ 
$1.0 < |y| < 1.5$          &    36/32       &       &   41/32  \\ 
$1.5 < |y| < 2.0$          &    43/28       &       &   60/28  \\ 
$2.0 < |y| < 2.5$          &    15/21       &       &   16/21  \\ 
$2.5 < |y| < 3.0$          &    4/18       &       &    5/18 \\
\hline  
low-$p\mathrm{_T}$ data                             &                &       &         \\
$0.0 < |y| < 0.5$          &           &   6/9    &   9/9  \\ 
$0.5 < |y| < 1.0$          &           &  18/9   &   45/9  \\ 
$1.0 < |y| < 1.5$          &           & 12/9       &   16/9  \\ 
$1.5 < |y| < 2.0$          &           & 11/9  &  12/9   \\ 
 $2.0 < |y| < 2.5$          &           & 12/9       &  11/9   \\ 
$2.5 < |y| < 3.0$          &           & 12/9  &   14/9  \\
 $3.2 < |y| < 4.7$          &           & 11/9   &   11/9  \\
\hline 
\end{tabular}
\end{center}
\end{table*}

Here the large model uncertainty comes from the fit where the $Q^2_{min}$ value of the DIS HERA data was set to 5 GeV$^2$ and the large experimental uncertainty is connected with the larger statistical uncertainties of the low-$p\mathrm{_T}$ data set. 
In all fits with free $\alpha_s(M_Z^2)$, as expected, the largest uncertainty comes from investigating the scales. NNLO interpolation of grids for the measurement considered here have  been made publicly available very recently~\cite{britzger_2022}, after this analysis was finalized. As a next step the inclusion of NNLO corrections into analysis is planned. These corrections are expected to improve the description of the data, and relieve some of the tensions found in this analysis. 

The total $\chi^2$ per number of degrees of freedom (NDF) and the partial 
$\chi^2$ per number of data points (NDP) for the data included in the fits described above are listed in Table~\ref{tab:1}.
Looking at the estimated values of $\alpha_s(M_Z^2)$ and the $\chi^2$ values in Table~\ref{tab:1}, we can see that there is a small tension between the low- and high-$p\mathrm{_T}$ jet data sets. 
The largest difference in the partial $\chi^2$ values for the fit with both low-$p\mathrm{_T}$ and high-$p\mathrm{_T}$ jet data sets can be observed in two $|\eta|$ bins: $1.5 < |\eta| < 2.0$ for the high-$p\mathrm{_T}$ jets and $0.5 < |\eta| < 1.0$ for the low-$p\mathrm{_T}$ data. 
For other $|\eta|$ bins, the partial $\chi^2$ values are rather similar. 
We have conducted a study where the parameterisations for the PDF fit was established for each of these jet data sets separately
This does not influence the $\alpha_s(M_Z^2)$ value much; there is a similar pattern of partial $\chi^2$ values. 

As a cross check, we also used different published parameterisation in QCD global fits to estimate $\alpha_s(M_Z^2)$. 
The HERAPDF2.0 parameterisation and the one used in the CMS QCD analysis of the high-$p\mathrm{_T}$ jet 
data~\cite{Khachatryan_2017} were taken. 
The HERAPDF2.0 parameterisation has two less parameters than the one used in this analysis, $D_{u_v}$ and $E_{\bar{d}}$. 
On the other hand, the parameterisation from Ref.~\cite{Khachatryan_2017} has three additional parameters: $E_g$,  $D_{d_v}$ and $D_{\bar{d}}$. 
The values of $\alpha_s(M_Z^2)$ resulting from these fits are summarised in Table~\ref{tab:2}. 
The shown uncertainties are experimental only. The values are consistent within the uncertainties. 
The largest deviations are observed for the parameterisation from Ref.~\cite{Khachatryan_2017}, 
however for our fit with both the high- and low-$p\mathrm{_T}$ data all three additional parameters are negative at the starting scale. 
This is not a physical behavior and should be avoided. In this particular case it resulted in the gluon distribution going slightly
below 0 around $x = 1$ and in strange wiggles at large $x$ for the sea distributions: total sea $\Sigma$, 
anti-down quark $\bar{d}$ and strange sea. 

\begin{table*}[ht]

\begin{center}
\caption{Values of $\alpha_s(M_Z^2)$ for fits with high- and low-$p\mathrm{_T}$ data for different parameterisation discussed in the text.}
\label{tab:2}
\begin{tabular}{c|ccc}
\hline
\hline
used data & low- and high-$p\mathrm{_T}$ & high-$p\mathrm{_T}$ &  low-$p\mathrm{_T}$\\
\hline
parameterisation: & & $\alpha_s(M_Z^2)$  (exp. uncert.)& \\
\hline
 this paper& $0.1199 \pm 0.0011$ & $0.1218 \pm 0.0021$ & $0.1178 \pm 0.0029$\\
\hline
 HERAPDF2.0~\cite{h12015combination}& $0.1211 \pm 0.0011$ & $0.1232 \pm 0.0013$ & $0.1177 \pm 0.0025$\\
\hline
 from~\cite{Khachatryan_2017} & $0.1176 \pm 0.0014$ & $0.1192 \pm 0.0015$ & $0.1201 \pm 0.0030$\\
\hline
\end{tabular}
\end{center}

\end{table*}

\section{QCD fits with fixed $\alpha_s(M_Z^2)$}
\label{S:4}

\begin{figure}[h]
\centering\includegraphics[width=0.45\linewidth]{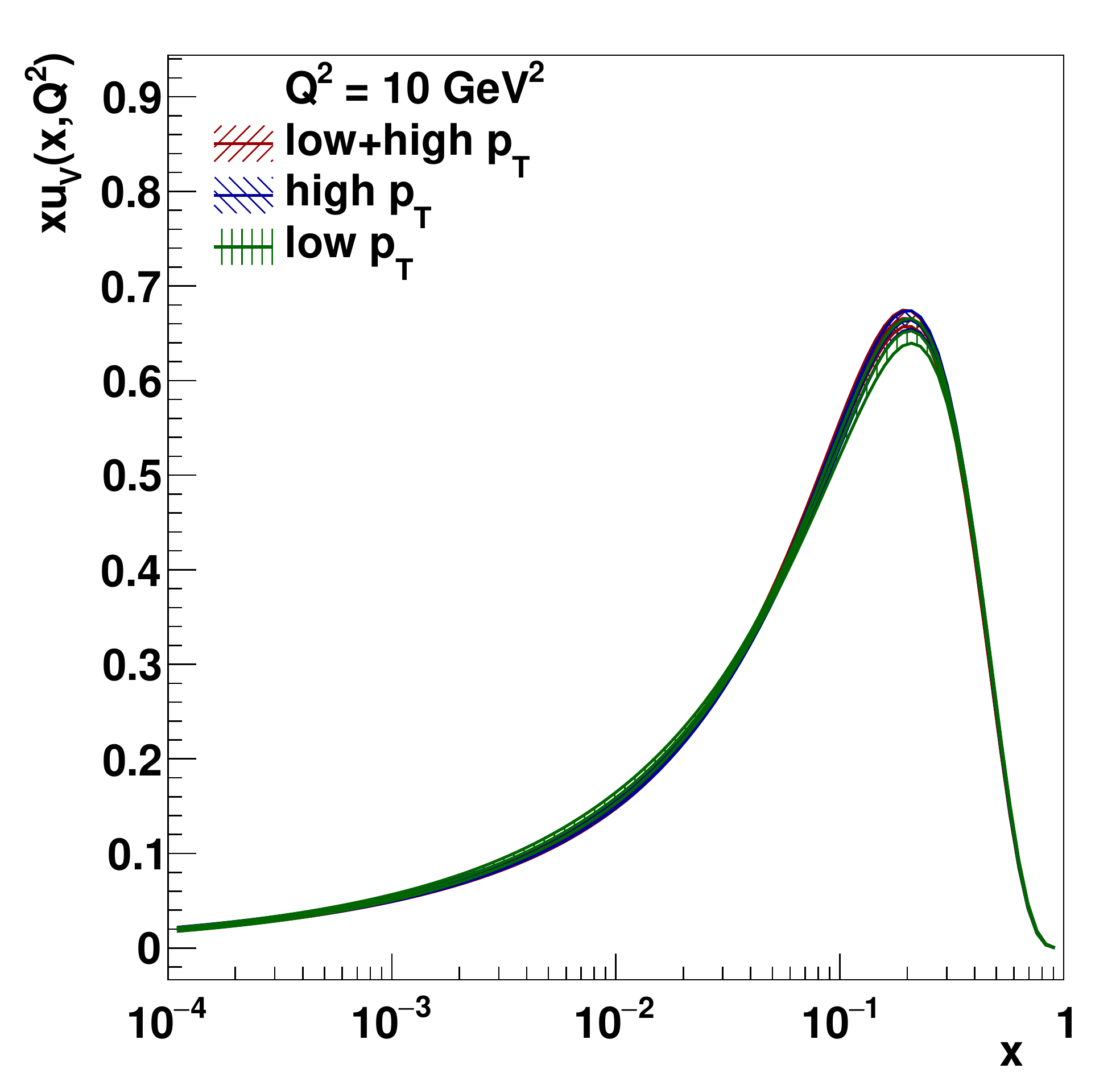}
\centering\includegraphics[width=0.45\linewidth]{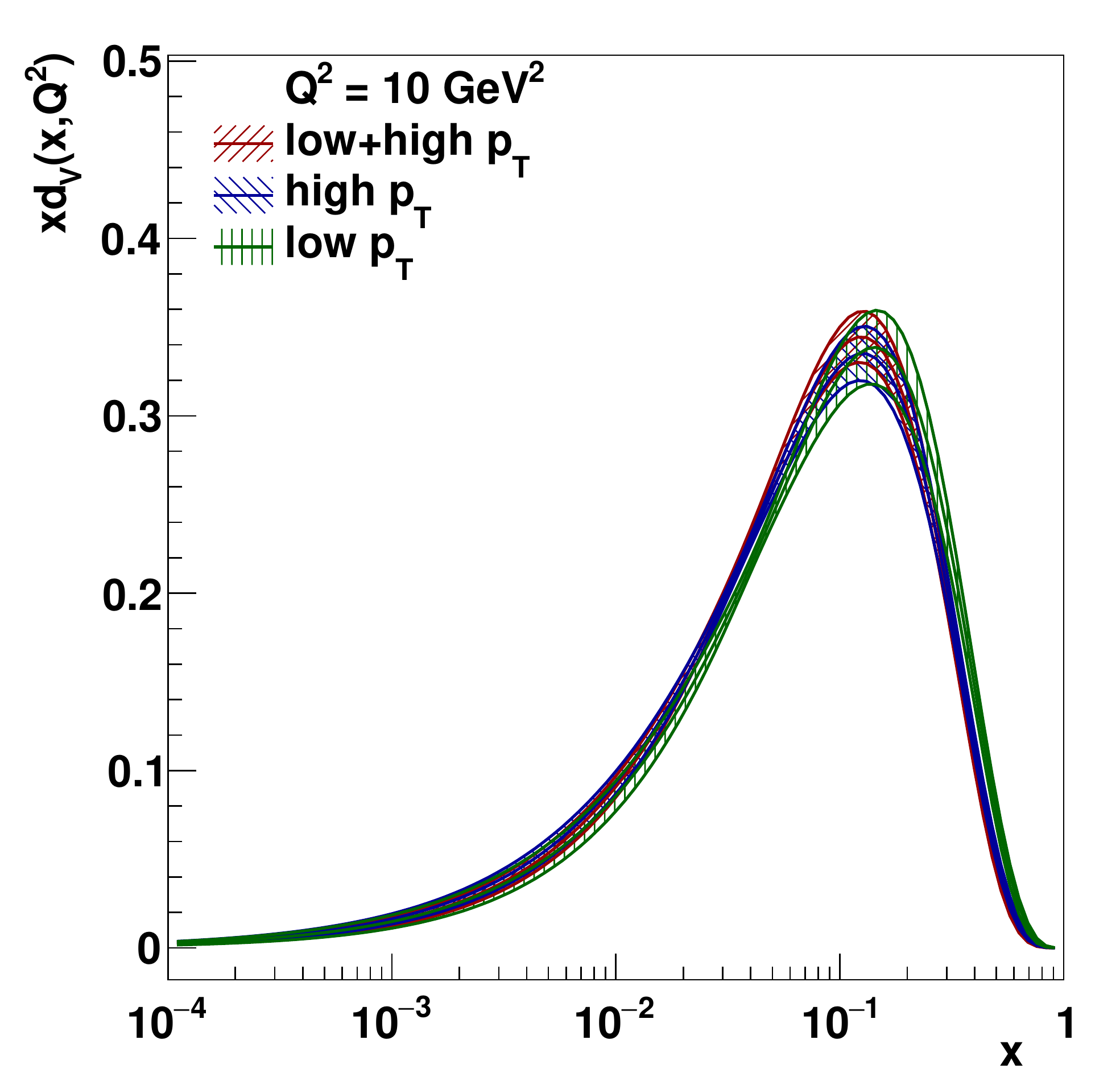}
\centering\includegraphics[width=0.45\linewidth]{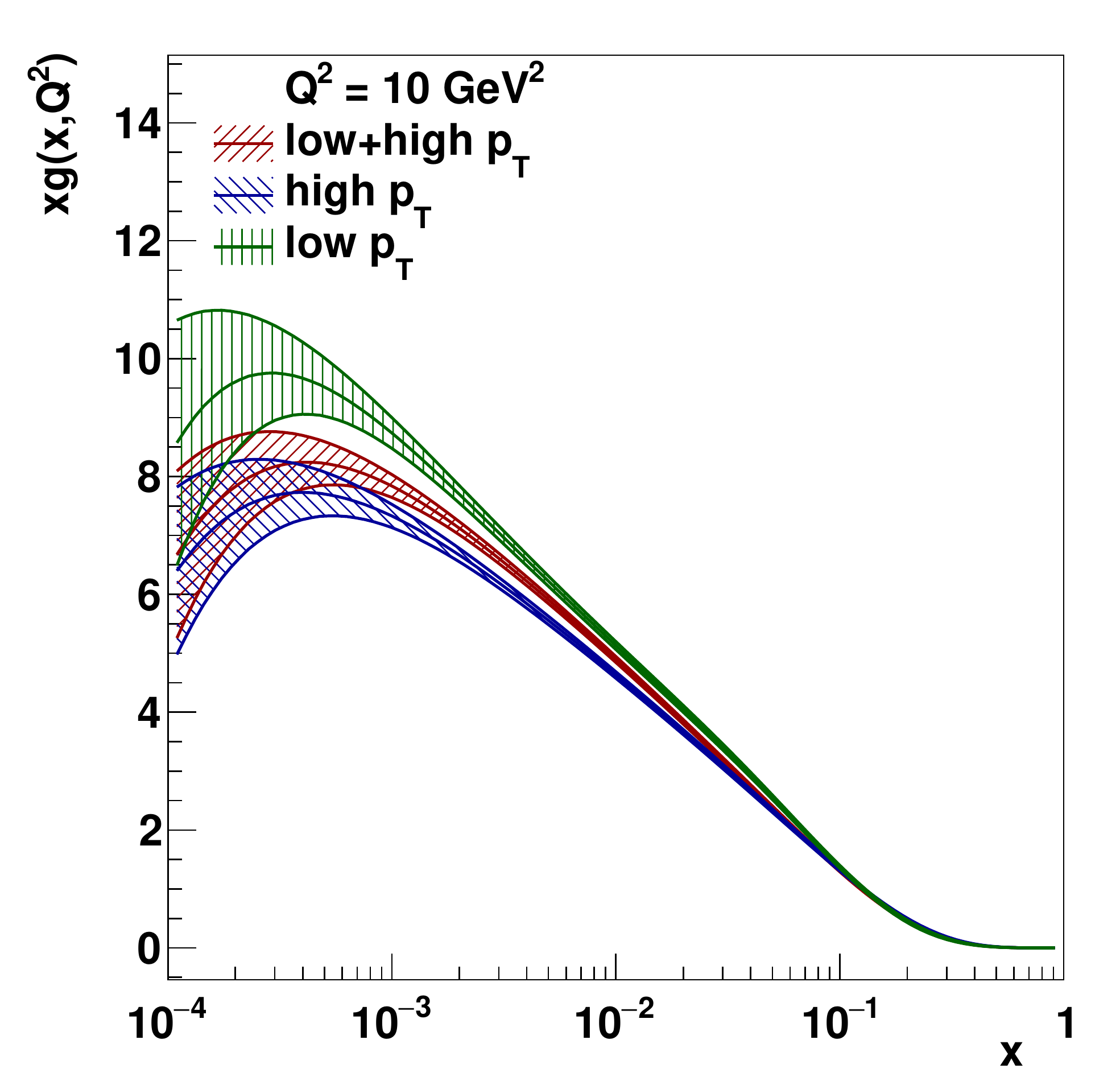}
\centering\includegraphics[width=0.45\linewidth]{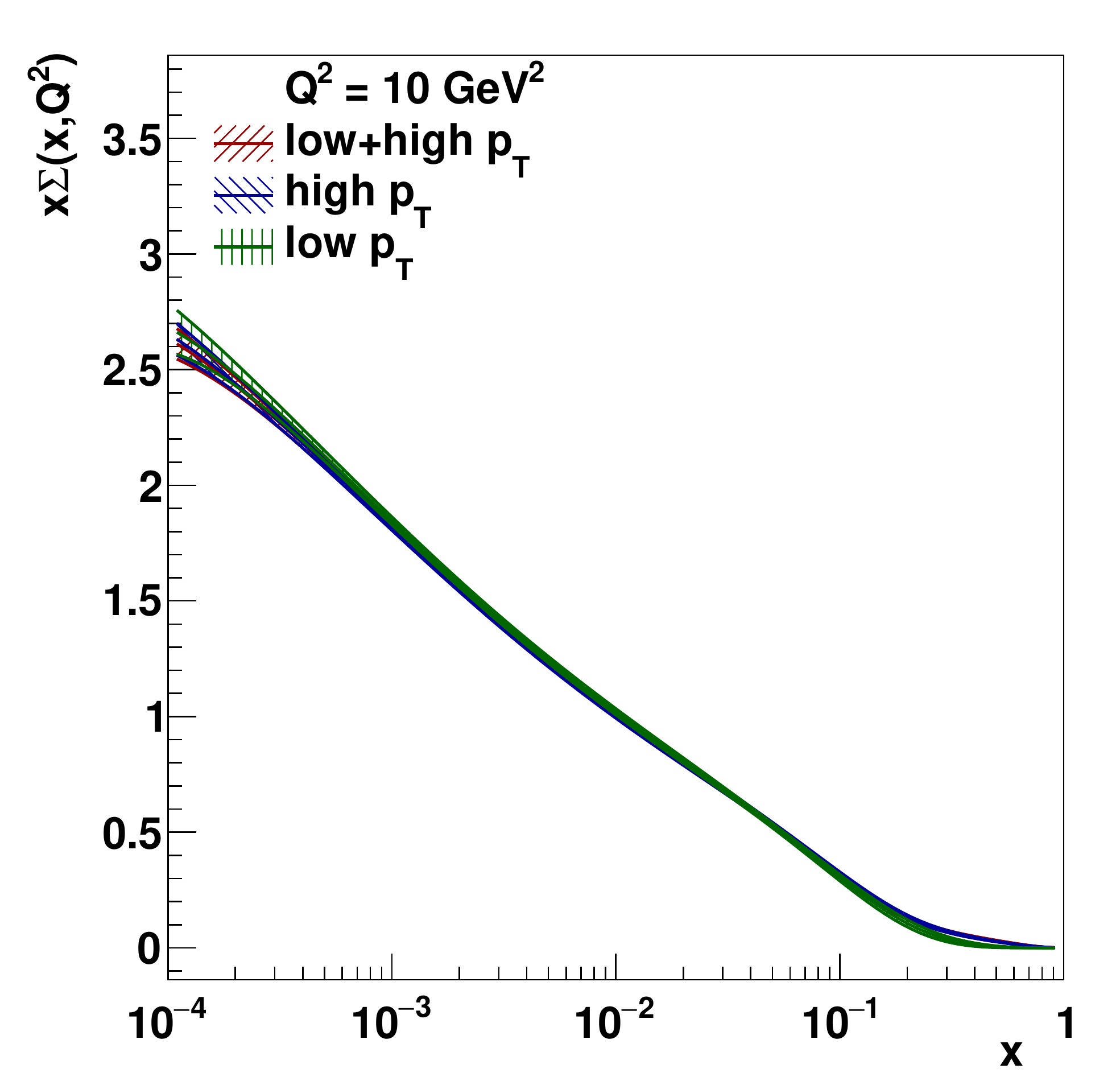}
\caption{Comparison of the valence quarks, gluon and sea distributions for the fits with both low- and high-$p\mathrm{_T}$ data,  high-$p\mathrm{_T}$ data only and low-$p\mathrm{_T}$ data only, at the scale of $10$~GeV$^{2}$.}
\label{fig-pdfs-10}
\end{figure}

Since the $\alpha_s(M_Z^2)$ values fitted to the separate data sets are different, we also performed the PDF fits including separately low- and high--$p\mathrm{_T}$ jet data, 
with $\alpha_s(M_Z^2)$ set to the optimal fitted value. 
Figure~\ref{fig-pdfs-10} shows a comparison of valence quarks, gluon and sea distributions at the scale of 10 GeV$^2$ for the three fits described above. 
The uncertainties are experimental only. 
The valence quarks and the sea are very well compatible within the uncertainties. The difference in the gluon distribution comes from the difference in $\alpha_s(M_Z^2)$.

\begin{figure}[h]
\centering\includegraphics[width=0.5\linewidth]{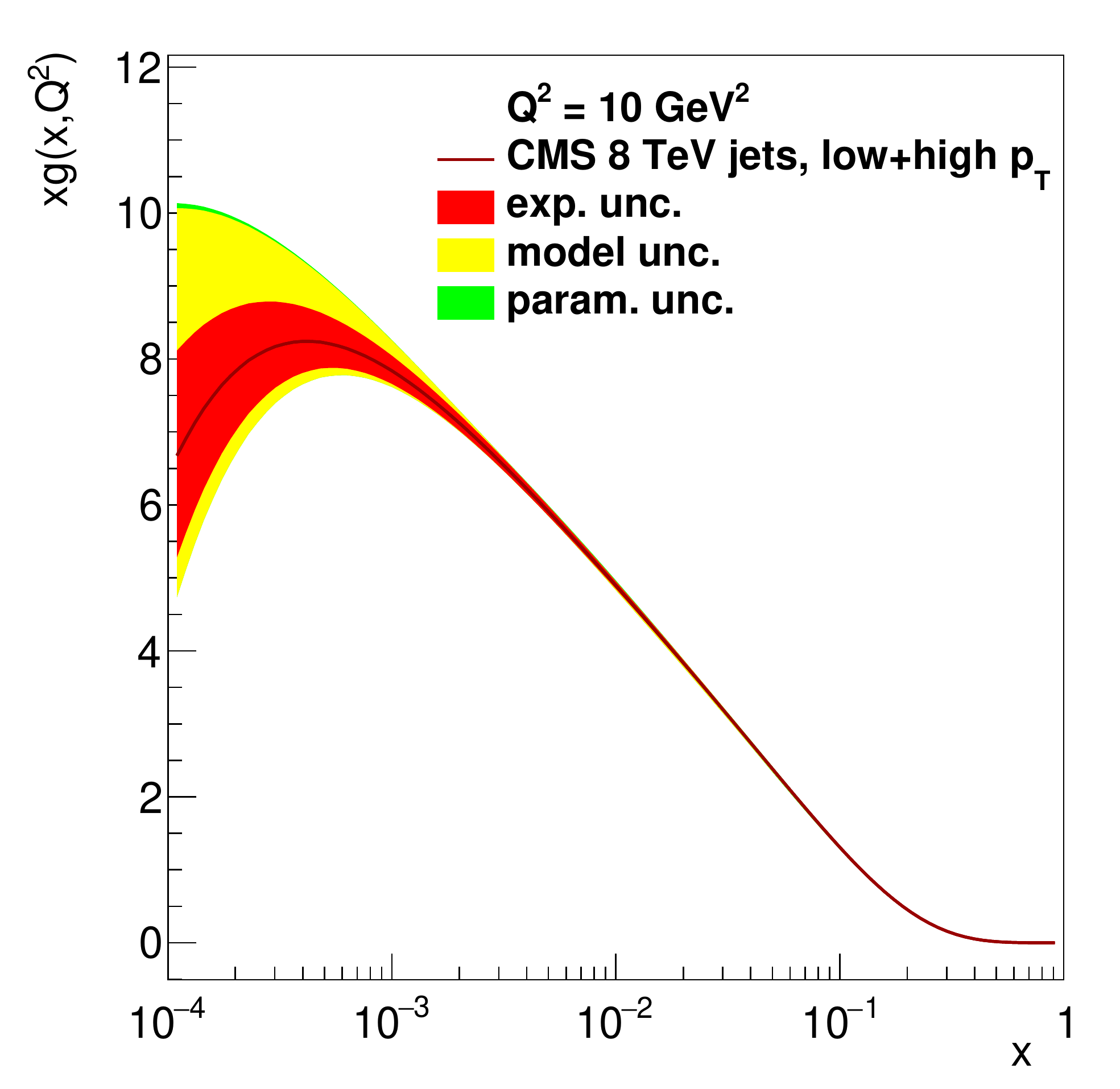}
\caption{Gluon distribution for the fit with both low- and high-$p_{\mathrm{T}}$ data at the scale of $0$~GeV$^{2}$.}
\label{fig-gluon-low+high}
\end{figure}

\begin{figure}[h]
\centering\includegraphics[width=0.45\linewidth]{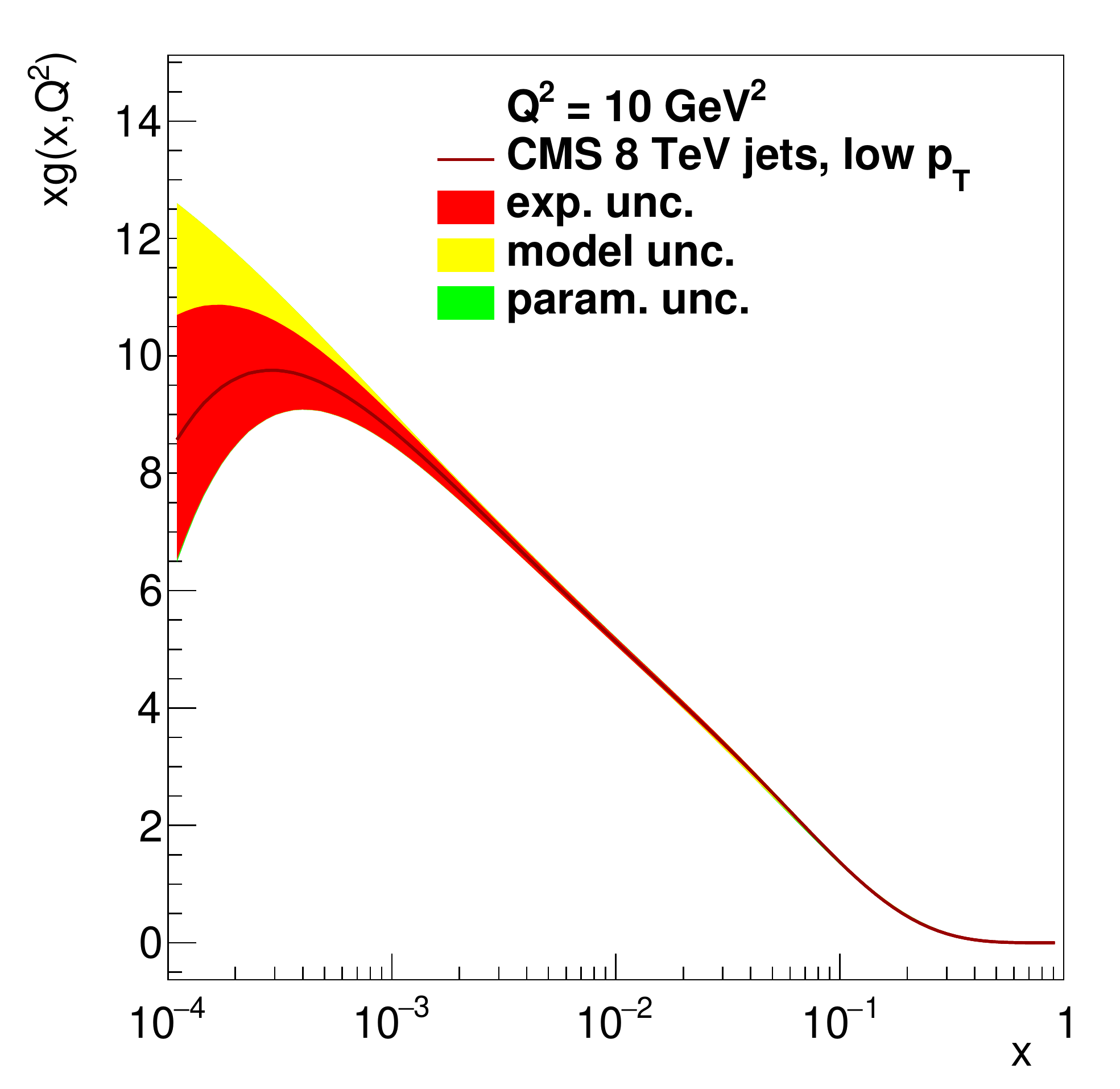}
\centering\includegraphics[width=0.45\linewidth]{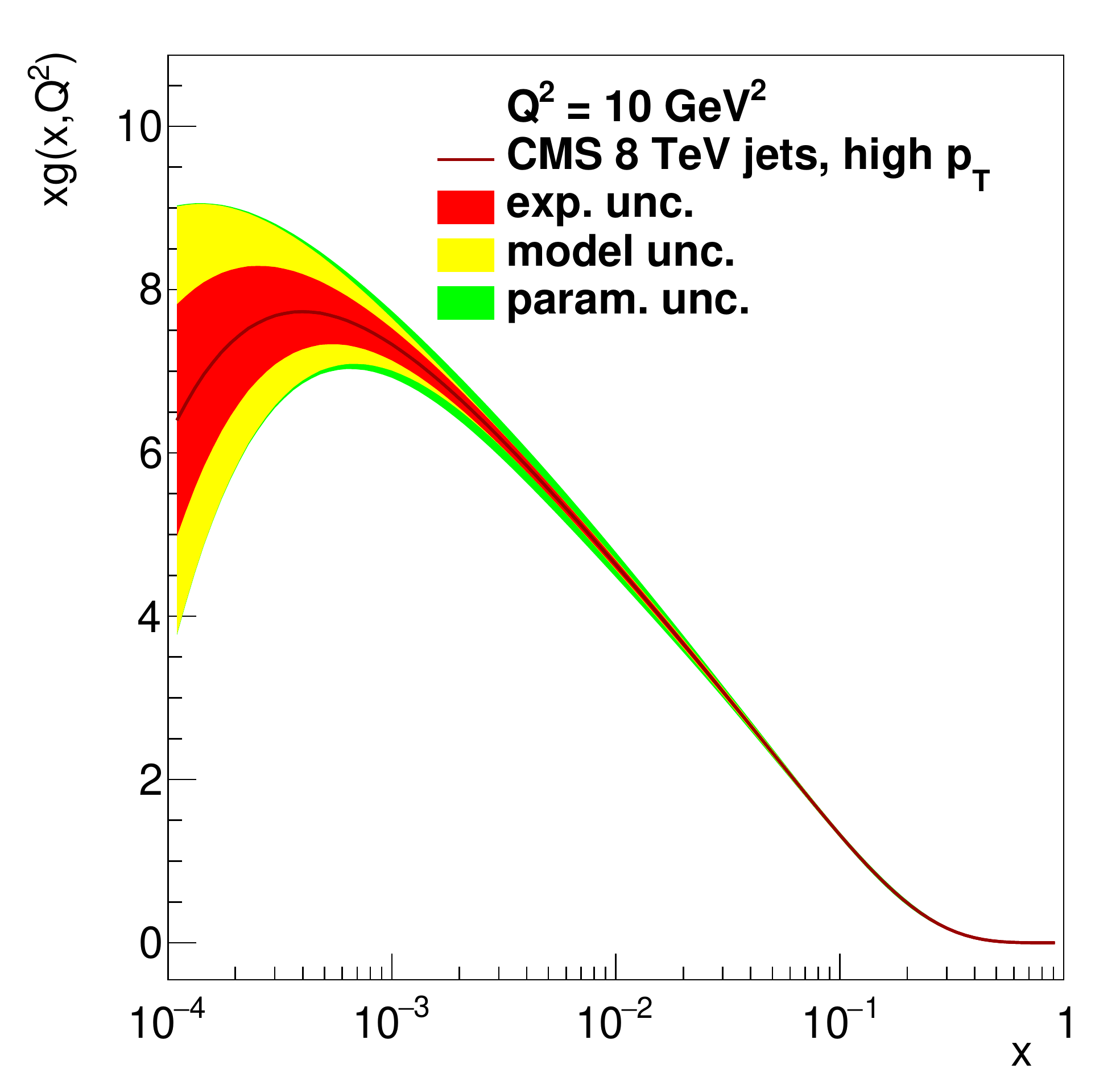}
\caption{Gluon distribution for the fit with low- and high-$p\mathrm{_T}$ data only at the scale of $10$~GeV$^{2}$.}
\label{fig-gluon-low-high}
\end{figure}

Figure~\ref{fig-gluon-low+high} shows the gluon distributions at the scale of 10 $\mathrm{GeV^{2}}$, with the total uncertainty, for the PDF fits with both low- and high-$p\mathrm{_T}$ jet data included. 
Figure~\ref{fig-gluon-low-high} shows the same distribution for the fits with only low-$p\mathrm{_T}$ and only high-$p\mathrm{_T}$ jet data included.
The colors indicate the experimental (red), model (yellow) and parameterisation (green) uncertainties. The dominant uncertainties are the experimental and model ones.
There is not much difference in the experimental, model, and paramaterisation uncertainties of the fits.

\subsection{Impact on the accuracy of the gluon distribution determination}
\label{ss:comparison}

\begin{figure}[h]
\centering\includegraphics[width=0.45\linewidth]{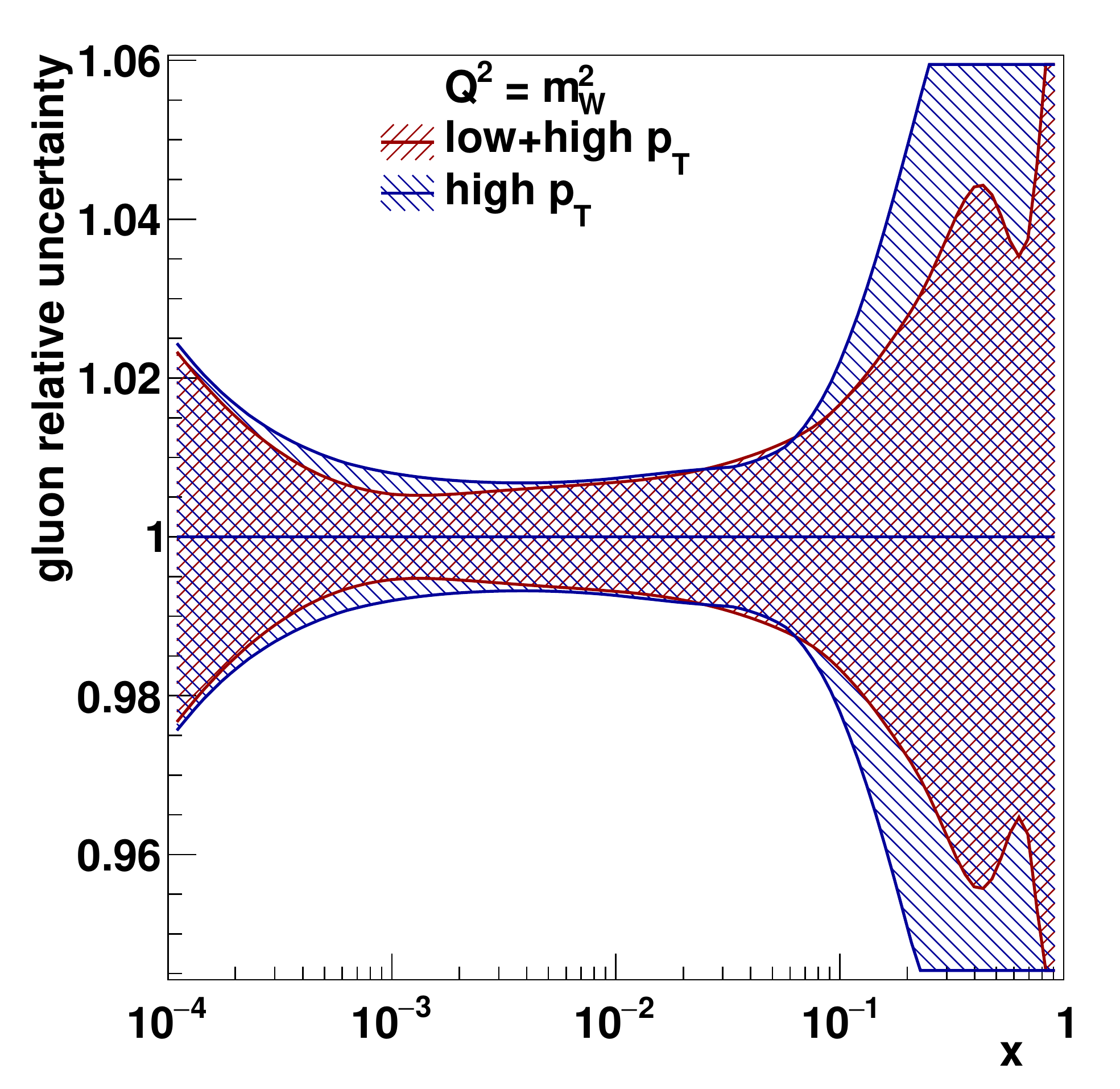}
\centering\includegraphics[width=0.45\linewidth]{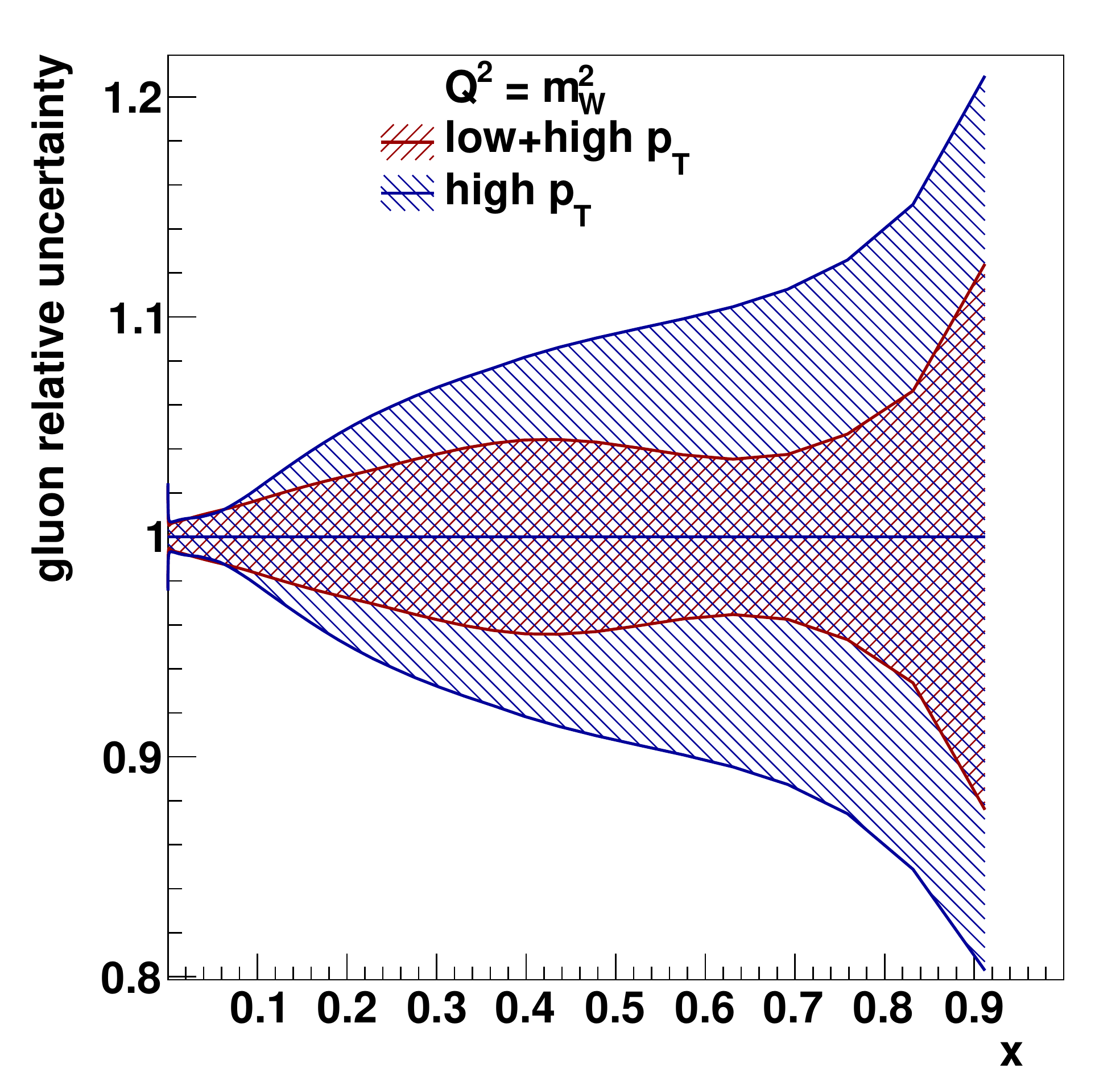}
\caption{Comparison of the relative total uncertainties for the gluon distribution at linear (left) and logarithmic (right) scale 
for $x$ with and without the low-$p\mathrm{_T}$ jets data for the scale of $m^2_W$.}
\label{comparison}
\end{figure}

To assess the impact of the low-$p\mathrm{_T}$ jet data on the gluon distribution, 
the distributions and uncertainties of two PDF fits are compared,  with and without these data,
using  the  same  parameterisation  described  in
Eqs.~\ref{eq:xgpar}-\ref{eq:xdbarpar} and for fixed $\alpha_s(M_Z^2) = 0.1199$.
The central values of all parton distributions in those two fits are consistent within experimental uncertainties.  
The relative total uncertainties for these two QCD fits are compared for the gluon PDF in Fig.~\ref{comparison} 
for the scale of $m^2_W$, displayed for the logarithmic and linear $x$ scale, 
the latter to enhance the high-$x$ region. 
There is a small reduction of the uncertainties for the distribution from high+low-$p\mathrm{_T}$ fit 
with respect to the one obtained without 
the low-$p\mathrm{_T}$ data around $x \approx 10^{-3}$. 
The reduction is more pronounced for large $x > 0.1$,
where one expects the low-$p\mathrm{_T}$ data to have the largest impact.
This is a good example of employing LHC jet data in a global QCD analysis.

\subsection{Comparison to data}
\label{SS:4-2}

Here we show how the predictions obtained using these fits compare to the experimental 8 TeV jet data used in the PDF analysis. The figures are done using xFitter and therefore the final predictions are represented by the dashed lines which correspond to the predictions after shifting them with the nuisance parameters that are the systematic uncertainties in the fit.

\begin{figure}[h]
\centering\includegraphics[width=0.45\linewidth]{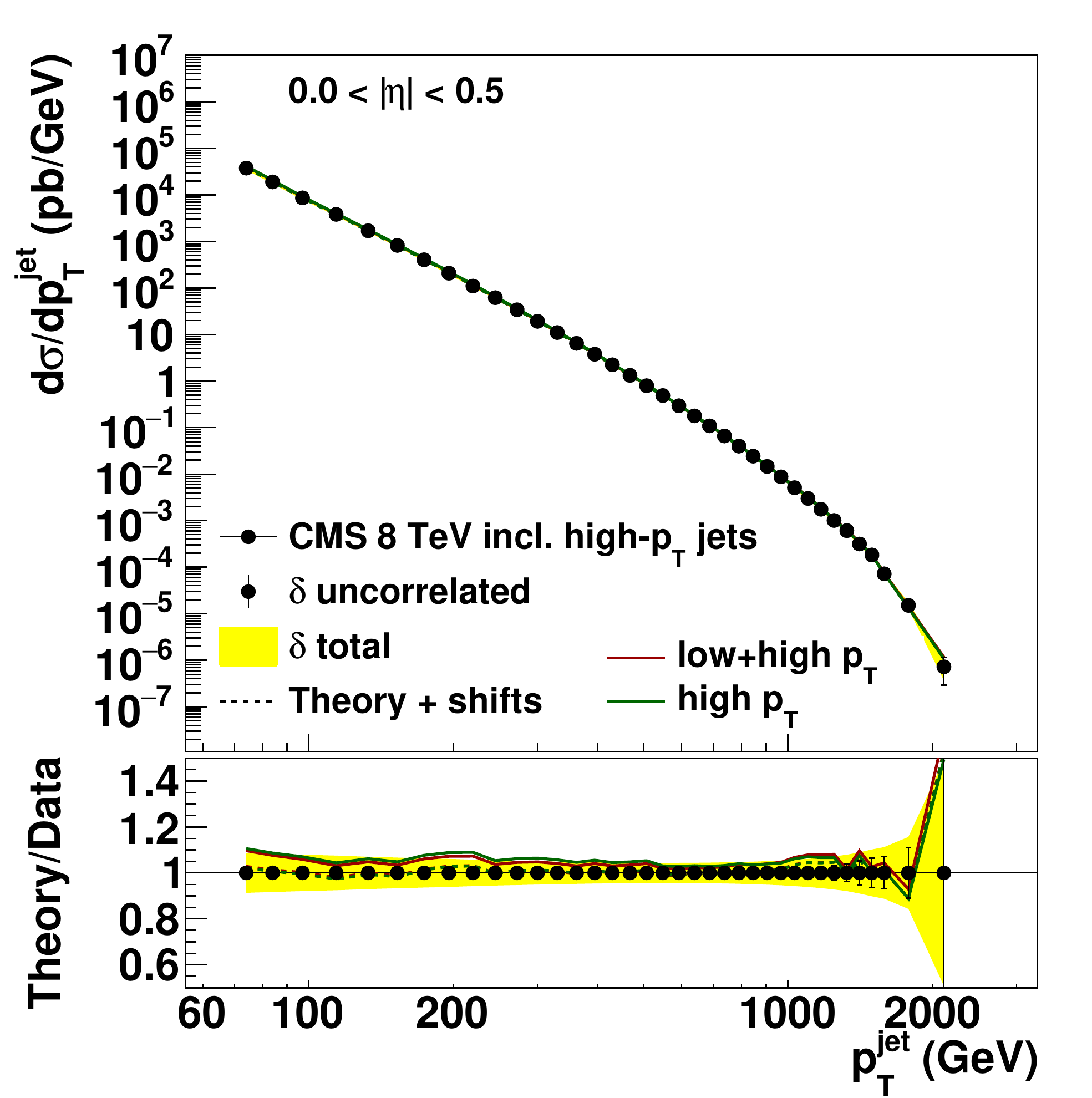}
\centering\includegraphics[width=0.45\linewidth]{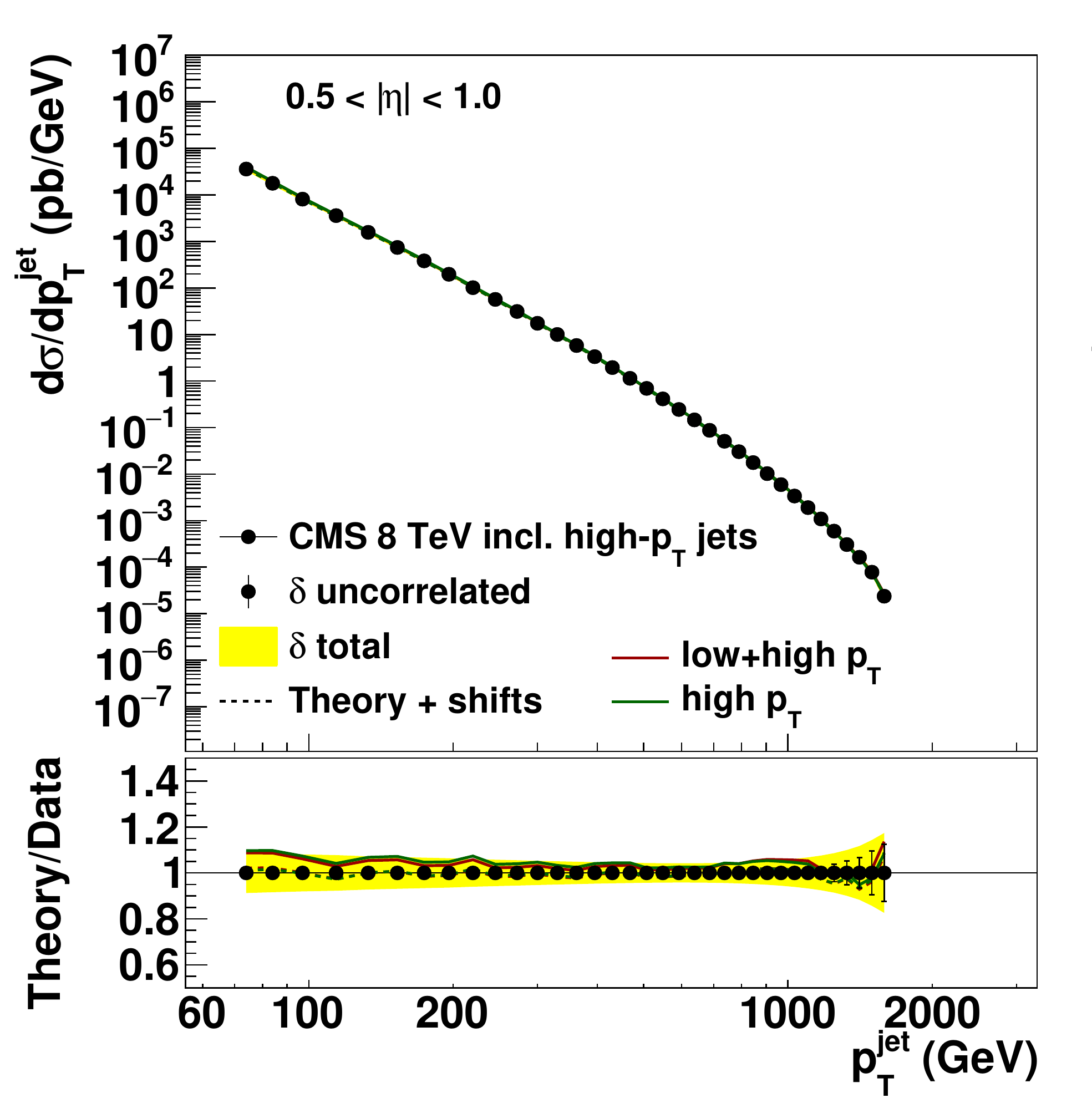}
\centering\includegraphics[width=0.45\linewidth]{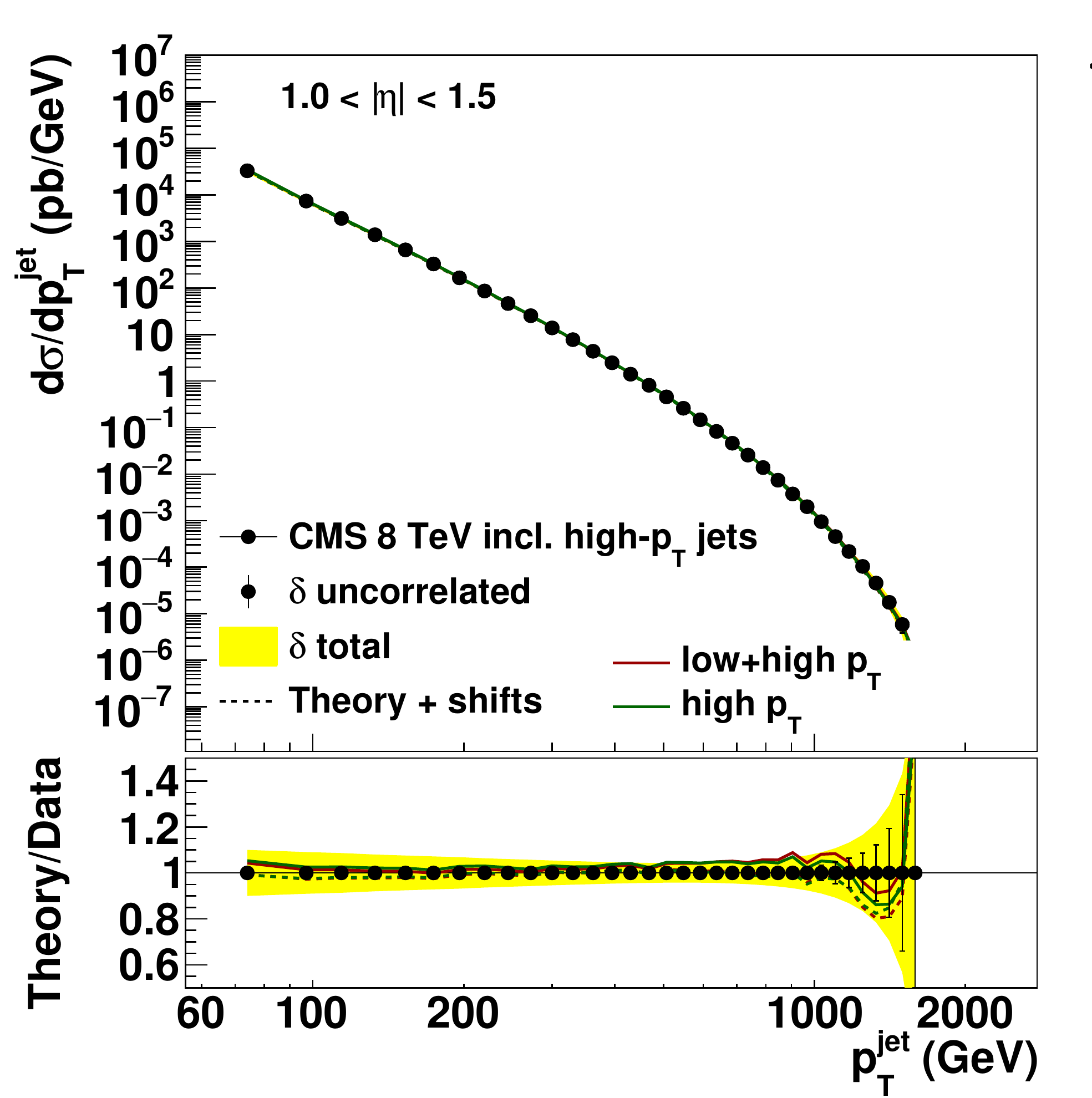}
\centering\includegraphics[width=0.45\linewidth]{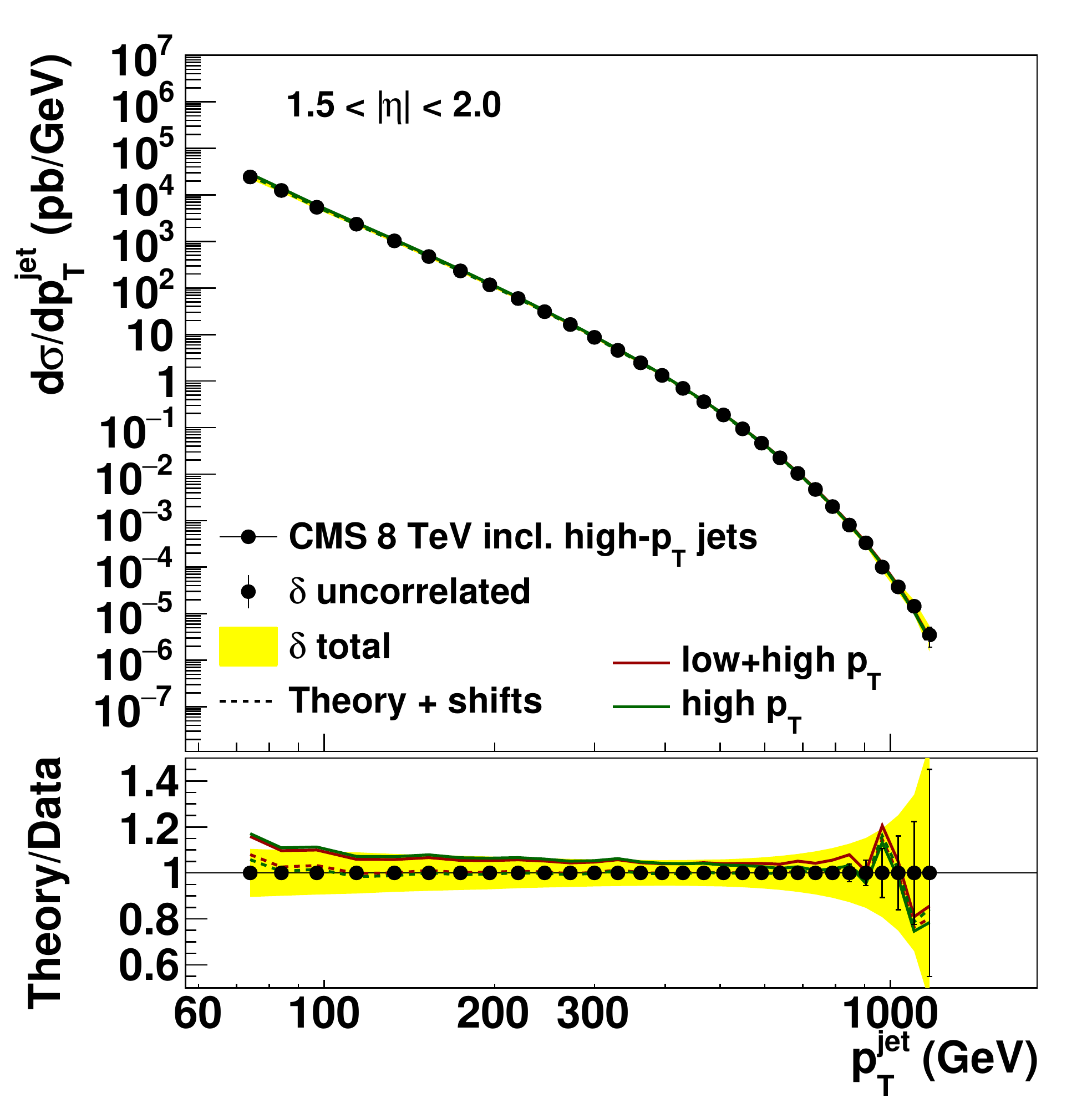}
\centering\includegraphics[width=0.45\linewidth]{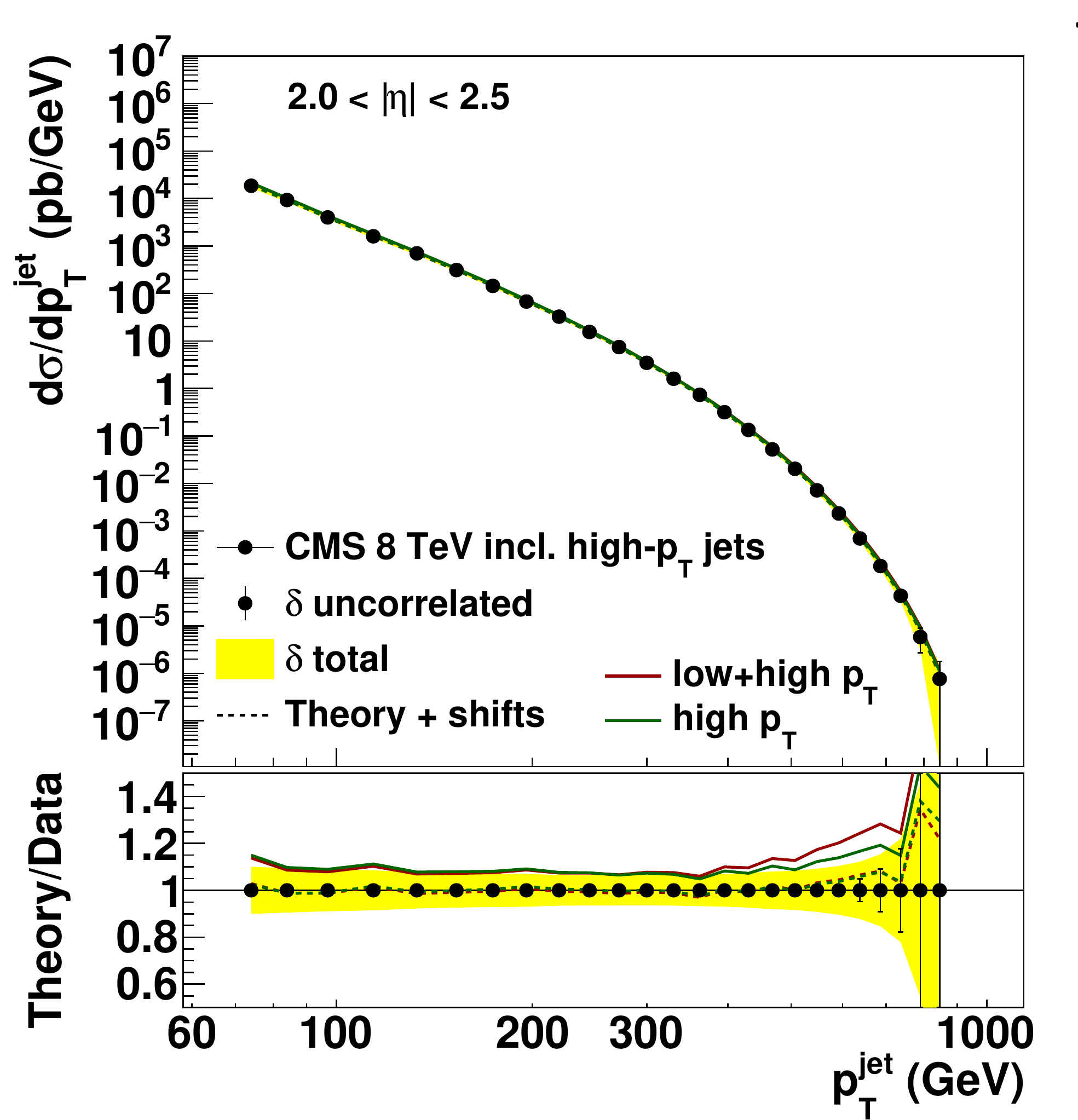}
\centering\includegraphics[width=0.45\linewidth]{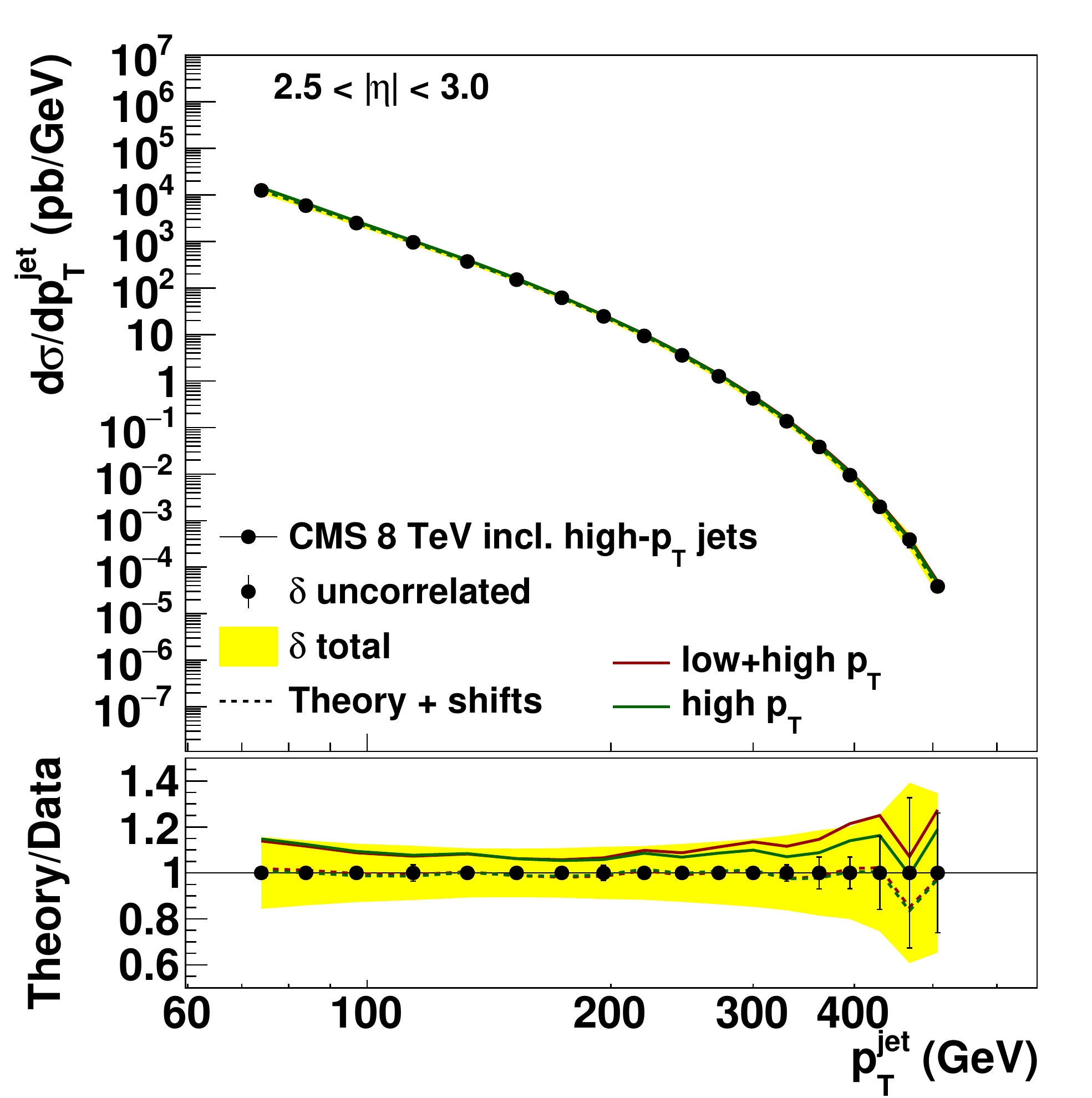}
\caption{Comparison of the high-$p\mathrm{_T}$ data with the theoretical predictions for two fits with $\alpha_s(M_Z^2)$ fixed to 0.1210, with high-$p\mathrm{_T}$ data and high+low-$p\mathrm{_T}$ data.}
\label{data1}
\end{figure}

\begin{figure}[h]
\centering\includegraphics[width=0.35\linewidth]{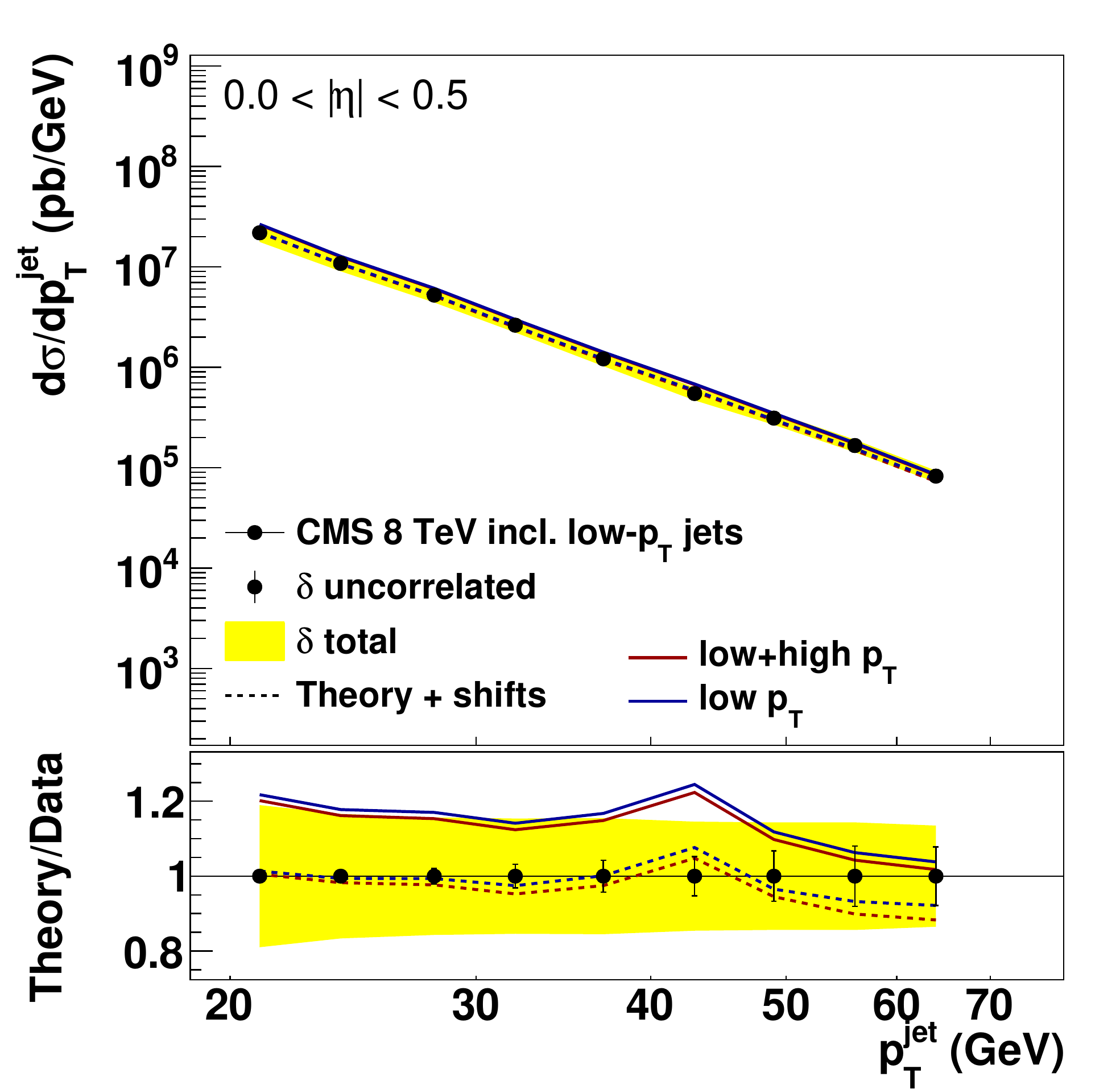}
\centering\includegraphics[width=0.35\linewidth]{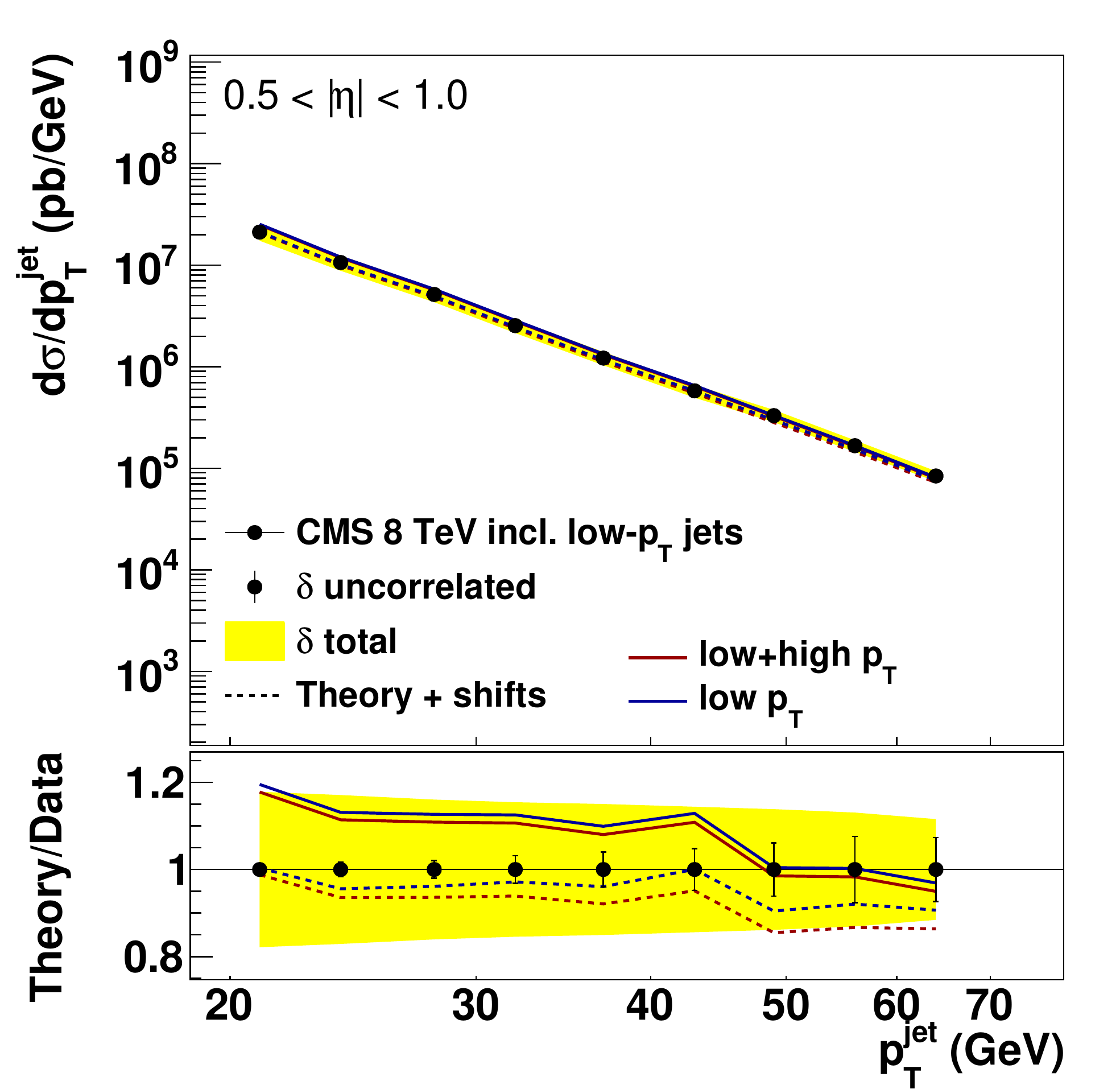}
\centering\includegraphics[width=0.35\linewidth]{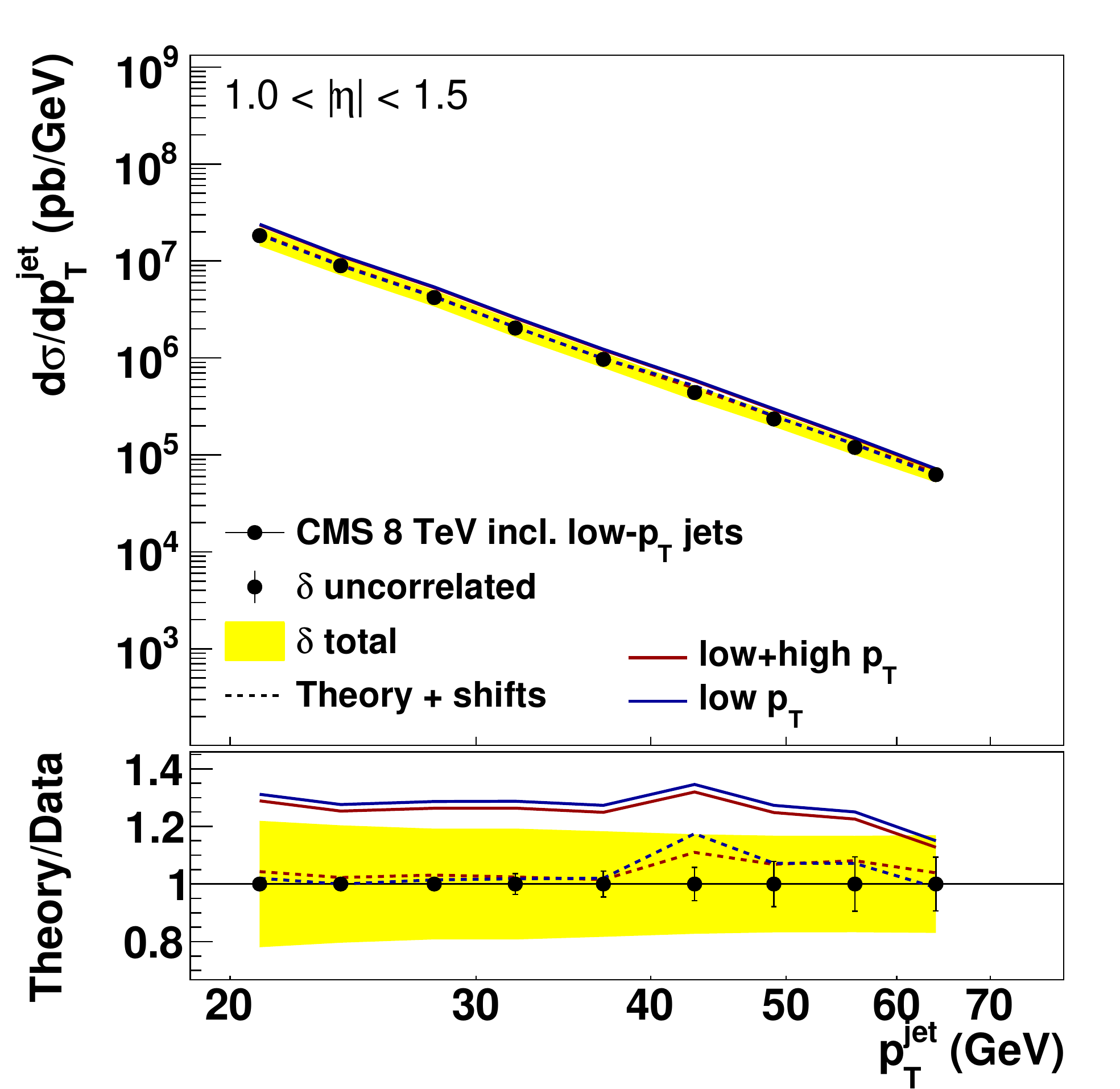}
\centering\includegraphics[width=0.35\linewidth]{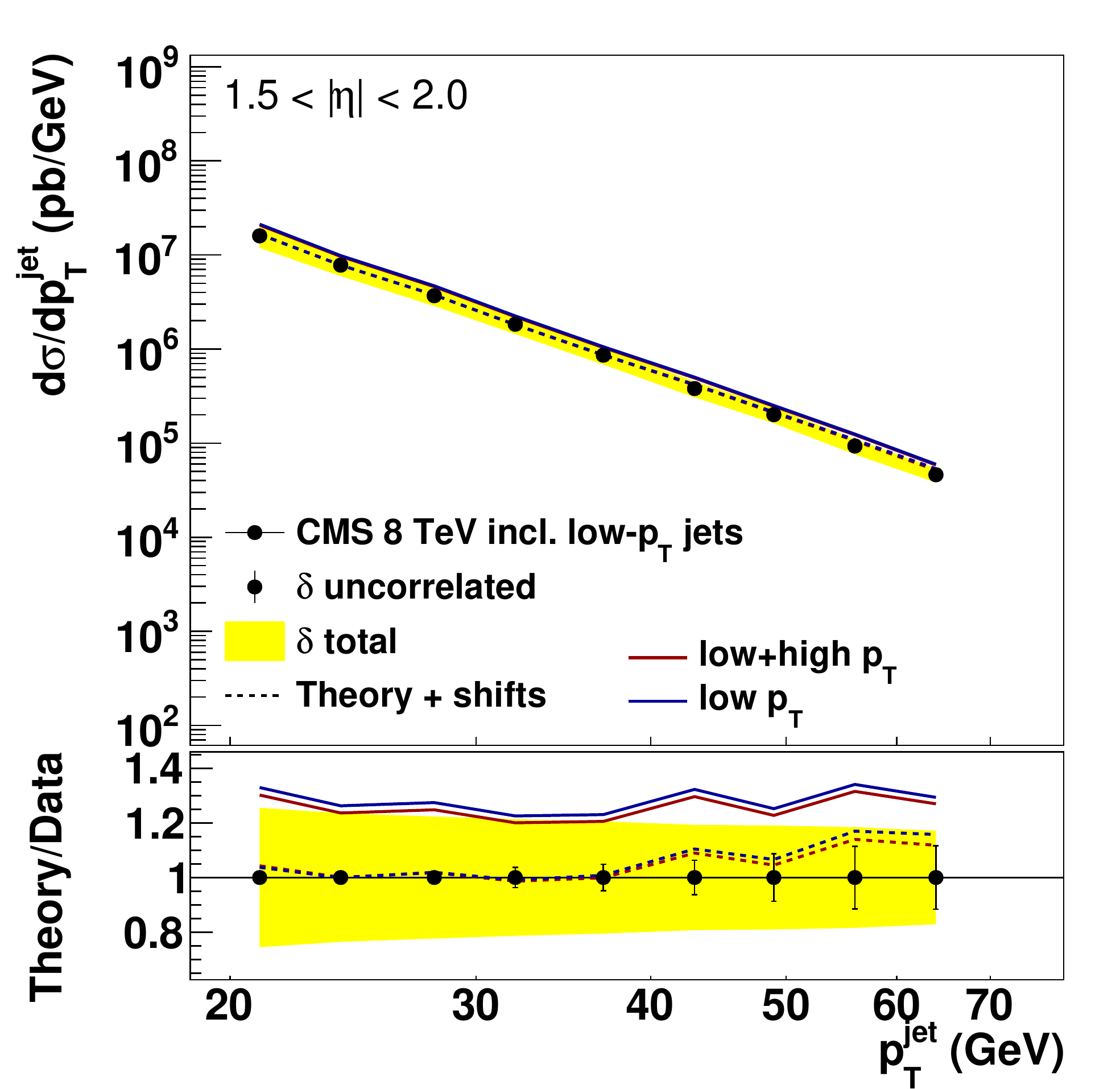}
\centering\includegraphics[width=0.35\linewidth]{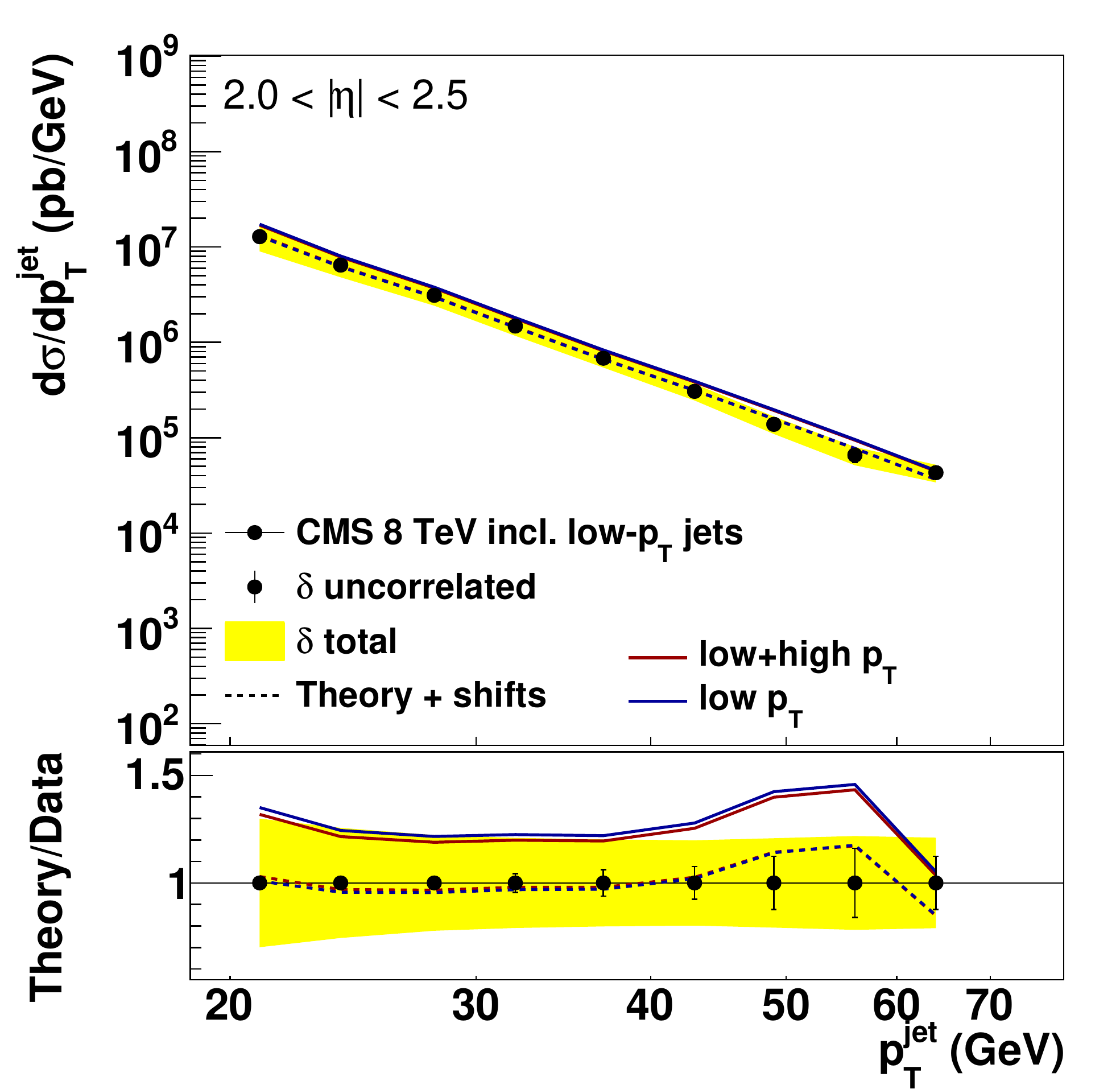}
\centering\includegraphics[width=0.35\linewidth]{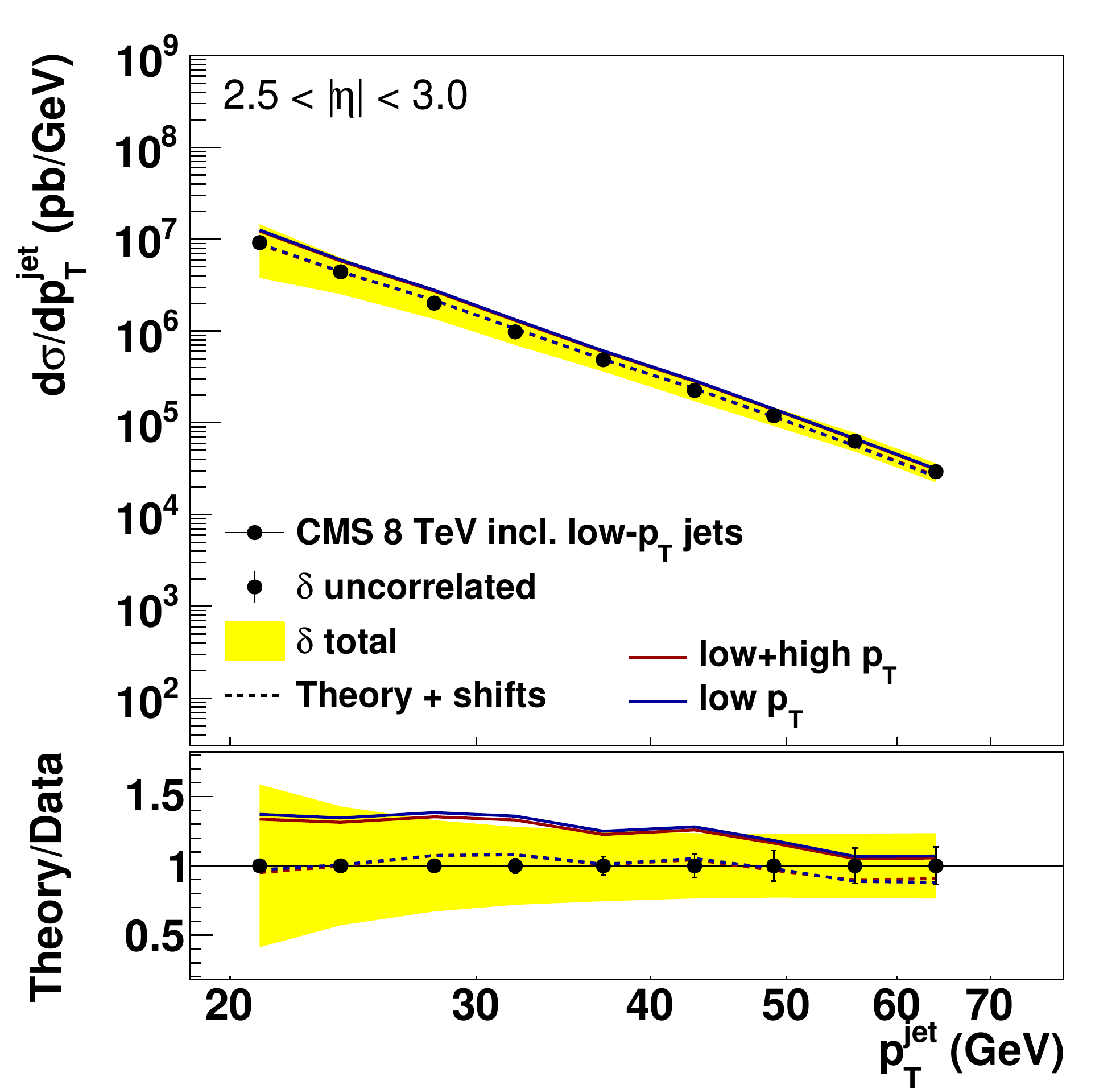}
\centering\includegraphics[width=0.35\linewidth]{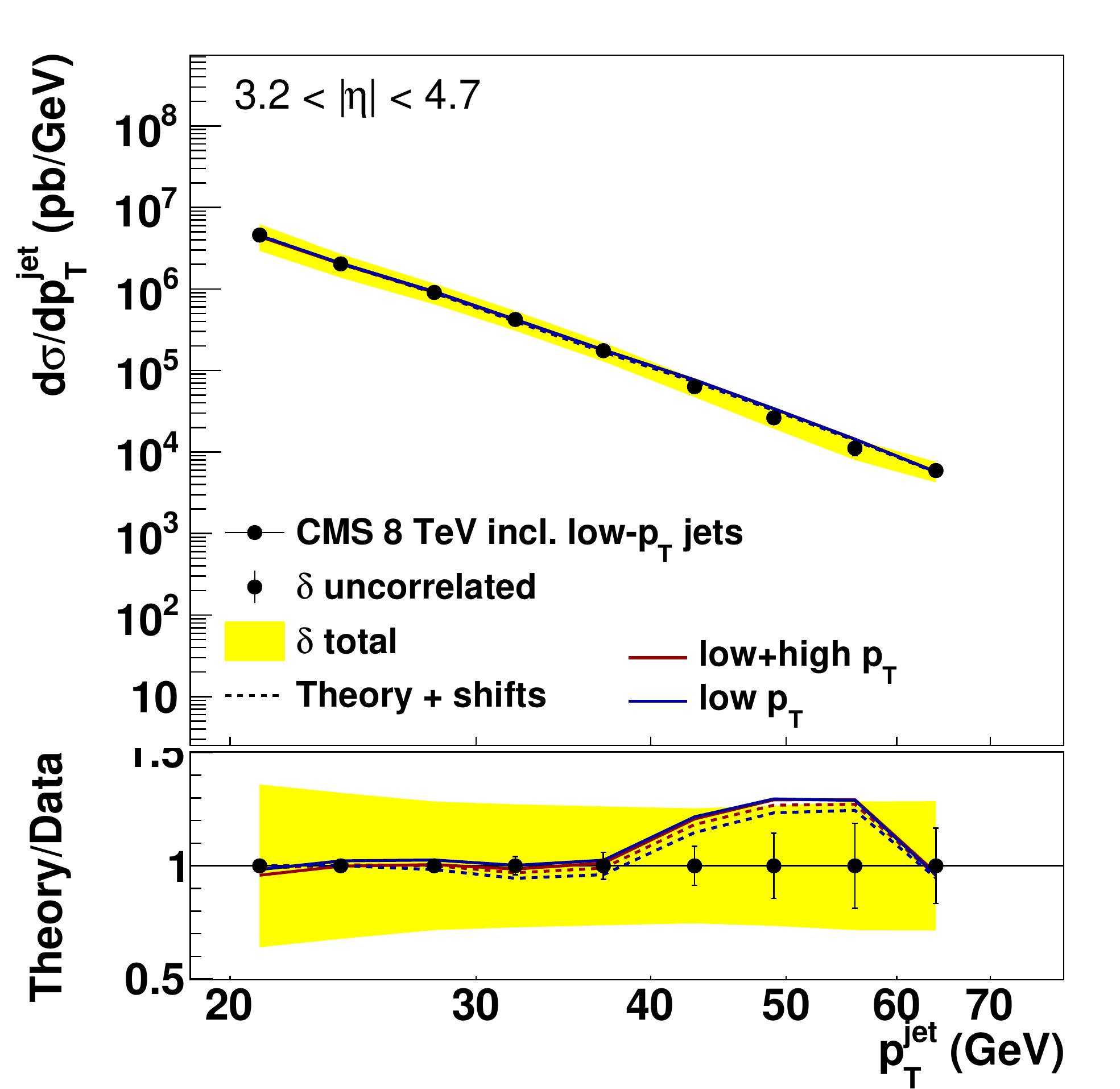}
\caption{Comparison of the low-$p_{\mathrm{T}}$ data with the theoretical predictions for two fits with $\alpha_s(M_Z^2)$ fixed to 0.1210, with low-$p_{\mathrm{T}}$ data and high+low-$p_{\mathrm{T}}$ data.}
\label{data2}
\end{figure}

Figure~\ref{data1} shows comparison of the high-$p\mathrm{_T}$ data with the theoretical predictions for two fits with $\alpha_s(M_Z^2)$ fixed to 0.1210, including high-$p\mathrm{_T}$ data only and high+low-$p\mathrm{_T}$ data. 
Figure~\ref{data2} shows comparison of the low-$p\mathrm{_T}$ data with the theoretical predictions for two fits with $\alpha_s(M_Z^2)$ fixed to 0.1210, including low-$p\mathrm{_T}$ data only and high+low-$p\mathrm{_T}$ data.
The overall description is reasonable. Some differences can be observed for some $\eta$ bins between the quality of both fits - as already seen from Table~\ref{tab:1}.

\section{Summary}

In the analysis described in this paper we have extracted the proton parton distribution functions (PDFs) at next-to leading order in perturbative quantum chromodynamics using the HERA inclusive cross sections and the CMS inclusive differential jet cross section data measured at center-of-mass energy of 8 TeV. 
The jet cross sections included the measurements at high and low jet $p_T$, and at forward rapidities.
A simultaneous fit of proton PDFs and the strong coupling $\alpha_s(M_Z^2)$ was performed using the xFitter open source platform, as well as a global QCD fits with $\alpha_s(M_Z^2)$ values fixed. 
The experimental, model and parameterisation, as well as the  theoretical scale uncertainties were evaluated. 
The jet data were compared to the theory predictions and they agree well. 
A small tension between the low- and high-$p\mathrm{_T}$ jet data sets was observed.
To asses the impact of low-$p_T$ jet measurements on the determination of the gluon distribution, 
PDF fits obtained with both low- and high-$p\mathrm{_T}$ data and with only 
high-$p\mathrm{_T}$ data were compared. 
In the fit including the low-$p\mathrm{_T}$ data a sizable reduction of the uncertainties for the gluon distribution was observed, in particular in the large-$x$ region $x > 0.1$.  Including the newly available NNLO corrections into this analysis is expected to improve the description of the data, and relieve some of the observed tensions between the data sets.

\label{S:5}

\clearpage




\end{document}